\documentclass[structabstract]{aa}  
\usepackage{graphicx}
\usepackage{color}
\usepackage{epsfig}
\usepackage{supertabular}
\usepackage{longtable,lscape}
\usepackage{subfigure}
\usepackage[varg]{txfonts}
\usepackage{natbib}
\bibpunct{(}{)}{;}{a}{}{,}
\def\mpc{h^{-1} {\rm{Mpc}}}
\def\kms {\rm{km~s^{-1}}}

\def\arcsec{''}
\def\arcmin{'}
\def\Msol {M_\odot}

\begin{document}
 \title{Galaxy properties in clusters: Dependence on the environment and the cluster identification techniques}


   \author{V. Coenda\inst{1,2} \and H. Muriel\inst{1,2}}


   \institute{Instituto de Astronom\'ia
  Te\'orica y Experimental IATE, CONICET, Laprida 922, X5000BGR, C\'ordoba, Argentina 
         \and Observatorio Astron\'omico, Universidad Nacional de C\'ordoba, Laprida 854, X5000BGR, C\'ordoba, Argentina\\
         \email{vcoenda@mail.oac.uncor.edu; hernan@.mail.oac.uncor.edu}
              }

   \date{}
   
 
\abstract
   {} 
   {We investigate the dependence of several galaxy properties on the environment and cluster 
identification techniques.}
   {We select clusters of galaxies from two catalogues based on the SDSS: the ROSAT-SDSS Galaxy Cluster 
Survey, which is an X-ray selected cluster sample and the MaxBCG Catalogue, in which clusters are optically 
selected. Based on a volume limited sample of galaxies drawn from the spectroscopic DR5 SDSS, we constructed sub-samples of 
clusters of galaxies with more than ten members. Scaling relations as well as segregation of galaxy 
properties as a function of the normalized clustocentric radii are analyzed. The properties of galaxies in clusters are compared with those of field galaxies.}
   {Galaxies in X-ray and MaxBCG selected clusters show similar size-luminosity relations. At equal 
luminosity, late type galaxies in the field have sizes smaller than cluster 
galaxies of the same morphological type.
 
The Faber-Jackson relation for early-type galaxies in clusters is also the same for X-ray selected 
and MaxBCG clusters. We found clear differences between the dynamical properties of galaxies in 
clusters, the brightest cluster galaxies ($BCG_s$) and field  galaxies.

Using several criteria to classify galaxies into morphological types, we reproduce the well 
know morphological segregation. The correlation is up to $r/r_{200} \sim 1$. For the whole range 
of clustocentric distances, X-ray selected clusters present a higher fraction of early type galaxies 
than MaxBCG clusters. We also found that bright galaxies preferentially inhabit the cluster centers. 
Median sizes of galaxies, such as the radius that enclose $50\%$ of Petrosian flux $r_{50}$, present a behaviour that also depends on the cluster selection criteria. 
For galaxies in X-ray selected clusters, median values of $r_{50}$ decrease as $r/r_{200}$ goes to zero, 
whereas the opposite is observed for galaxies in the MaxBCG clusters. These different behaviours are 
mainly due to early type galaxies.

The results are discussed in terms of the different processes that affect the evolution of 
galaxies in different environments.}
   {}


   \maketitle
%
\section{Introduction}
\label{sec:intro}

It is well known that galaxies show a wide range of morphologies, which can be appreciated 
in their properties such as colour, luminosity, size, star formation histories, etc, which 
imply that galaxies form and evolve through different mechanisms. 
There have been many studies about the distribution of galaxies with respect to their 
properties. For example, the luminosity function has been measured from various surveys of 
galaxies and clusters of galaxies and it is found to be well described by the Schechter 
function (\citealt{Schechter:1976}, \citealt{Loveday:1992}, \citealt{Folkes:1999}, 
\citealt{Mad:2002}, \citealt{Cross:2004}); the morphological types of galaxies are found 
to be correlated with the environment, clusters of galaxies being the best example 
(\citealt{Dressler:1980}, \citealt{Dressler:1997}, \citealt{Dominguez:2001}, \citealt{Coenda:2006}). 
It is also well known that different
galaxy properties are correlated to each other. Galaxy sizes are correlated with the luminosity 
and morphological type (\citealt{AP&B:1995}, \citealt{MH:2001}, \citealt{Coenda:2005}, 
\citealt{McI:2005}, \citealt{Trujillo:2006}), and have a distribution that may be described 
by a log-normal function (\citealt{Syer:1999}, \citealt{deJL:2000}, \citealt{Shen:2003}, 
\citealt{Ferguson:2004}). In order to constrain the galaxy formation models and the study of galaxy 
properties and their dependence on environment, we concentrate our studies on several 
scaling relations and galaxy segregation.

There are different scaling relations between photometric and structural parameters of 
galaxies, resulting in well known relations. One of the most established empirical scaling 
relations of disk galaxies is the Tully-Fisher relation (\citealt{TF:1977}), which consist 
of a correlation between luminosity and rotational velocity. The analogous relation for
spheroidal galaxies is the correlation between the velocity dispersion of bulges and luminosity, 
known as the Faber-Jackson relation: $L\propto \sigma^{\beta}$ (\citealt{FJ:1976}). Other scaling 
relations for spheroidal galaxies are: 
color-magnitude \citep{SV:1978b,SV:1978a}, color-velocity dispersion \citep{Bernardi:2005}, 
radius-luminosity \citep{SP:1990} and the Kormendy relation, which is a correlation between 
radius and surface brightness \citep{Kor:1977}. Some of these properties have been combined 
to define the so-called \emph{fundamental plane} that relates the velocity dispersion, the effective 
radius and the luminosity (\citealt{DD:1987}, \citealt{Dressler:1987}, \citealt{Bernardi:2003}). 
These empirical relations are closely related to the physical processes involved in the galaxy 
formation scenario and, therefore, are a fundamental tool to understand the formation and 
evolution of galaxies.  
These relations could depend on the environment where galaxies form and evolve, introducing 
departures from scatter in the different scaling relations. Several authors have investigated 
the scaling relations of galaxies in clusters. \citet{Ziegler:1999}, \citet{LaBarbera:2004}, \citet{Bernardi:2007}, \citet{vonderlinden:2007}, \citet{Liu:2007} and 
\citet{Bildfell:2008}. \citet{vonderlinden:2007} found that brightest cluster galaxies
(BCG) have a higher fraction of dark matter and consequently larger radii and higher velocity 
dispersions that non-BCG galaxies. \citet{Malumuth:1981,Malumuth:1985}, \citet{OH:1991} and more recently \citet{Bernardi:2007} found that BCG have 
smaller $\beta$ than the rest of the cluster members. \citet{Bernardi:2007} also analyze the 
dynamical mass ($M_{dyn} \propto r_{50} \sigma ^2$) and found that 
BCGs show a steeper relation than non-BCG galaxies. The majority of the works mentioned 
above do not make a suitable comparison with field galaxies, making it very difficult ti perform a complete 
comparison of the scaling relations of galaxies as a function of the environment.
More recently, \citet{Weinmann:2009} and \citet{Guo:2009} analyzed samples of galaxies in groups and found that, at fixed stellar mass, the sizes of central 
and satellite early type galaxies are similar, while \citet{Bernardi:2009}, analyzing galaxies in clusters, reported a difference at high luminosities/stellar masses between central and satellite galaxies.

It is well known that different types of galaxies show different spatial distributions 
(\citealt{Oemler:1974}, \citealt{MS:1977}, \citealt{Dressler:1980}). These differences 
can be estimated measuring the radial dependence of the galaxy 
properties as a function of the clutocentric distance.  The segregation of galaxies in clusters has been quantified splitting galaxies 
according to different properties, with
the morphology the property that has been most extensively studied, see for instance 
\citet{Dressler:1980}, \citet{Whitmore:1993}, \citet{Dominguez:2001}, \citet{Biviano:2002}, 
\citet{Driver:2003}, \citet{Coenda:2006}. \citet{Goto:2002} found that the morphological segregation 
tends to disappear for clustocentric distances larger than the virial radius. Recently, \citet{MCM:2008} found that 
$g-r$ color is the property of galaxies that best predict the normalized clutocentric distance
of galaxies among a set of galaxy properties. 
Similarly, \citet{Skibba:2008} used the galaxy sample of visual classified 
morphologies and found that much of the morphology-density relation is due to the relation between 
colour and density. The works that 
have studied the segregation in luminosity present some contradictory results. \citet{RT:1968}, 
\citet{Yepes:1991}, \citet{Lobo:1997}, \citet{Kashi:1998}, \cite{Coenda:2006} found results consistent 
with a segregation in luminosity. Nevertheless, \citet{Noonan:1961}, \citet{Bahcall:1973} and more 
recently \citet{Pracy:2005} found results consistent with a lack of segregation in luminosity. 
The segregation of other properties like color, star formation rate, gas content or structural 
parameters have been much less studied.

The identification of clusters of galaxies is a complex process whose outcome depends on 
the selection criteria. \citet{Popesso:2004} compiled X-ray clusters in the Sloan 
Digital Sky Survey (SDSS).
Using the same survey, \citet{Koester:2007:alg} applied a maxBCG algorithm to 
identified clusters. This technique is based on the color-magnitude relation of the early 
type galaxies and the properties of the BCGs. Because we are interested in comparing the galaxy 
properties of clusters selected with different criteria, the same analysis will be performed
on sub-samples of the catalogues compiled by \citet{Popesso:2004} and \citet{Koester:2007:alg}.

This paper is organised as follows: in section \ref{sec:sample} we describe the cluster and galaxy 
samples; in section \ref{sec:sr} we investigate several galaxy scaling relations; in section 
\ref{sec:gs} we analyze the galaxy segregation. We summarise and discuss our results in section 
\ref{sec:conc}.

\section{The sample}
\label{sec:sample}

\subsection{The cluster sample}
We use two catalogues of clusters of galaxies based on the Sloan Digital 
Sky Survey (SDSS; \citealt{York:2000}):
the ROSAT-SDSS Galaxy Cluster Survey of Popesso et al. (2004, hereafter P04), 
which is a X-ray selected cluster sample and the MaxBGC Catalogue of Koester et al. 
(2007b, hereafter K07), which is an optically selected cluster sample.  

The ROSAT-SDSS catalogue of \citet{Popesso:2004} comprises 114 galaxy clusters detected in the 
ROSAT All Sky Survey (RASS) lying in the area surveyed by the SDSS by February 
2003. This X-ray-selected catalogue includes clusters with masses from 
$10^{12.5}\cal{M}$$_{\odot}$ to $10^{15}\cal{M}$$_{\odot}$ in the redshift range 
$0.002\leq z \leq 0.45$. 
This catalogue provides X-ray properties of the clusters derived from the 
ROSAT data, parameters of the galaxy luminosity function and the 
luminosity of each cluster computed from SDSS data, and the radial 
distribution of the projected galaxy density around clusters.
Since the X-ray observations provide a robust method for identifying 
clusters, these clusters constitute a reliable source of information
to study galaxy properties and their correlation with the environment.

The optical MaxBGC catalogue provides sky locations, photometric redshift 
estimates and richness for 13823 clusters. Details of the 
selection algorithm and catalogue properties are published in 
\citet{Koester:2007,Koester:2007:alg}
The MaxBGC selection relies on the observation that the galaxy population 
of rich clusters is dominated by bright red galaxies tightly clustered 
in colour (the E/S0 ridgeline). Since these galaxies are old, passively 
evolving stellar populations, their $g-r$ colours closely reflect their 
redshifts. 
The brightest red galaxy, typically located at the peak of the galaxy density, 
defines the cluster centre. The K07 catalogue comprises galaxy clusters with 
velocity dispersions $\sigma \ge 400 \kms$ and redshifts $0.1\le z\le 0.3$. 
Each of these clusters contains between 10 and 190 E/S0 ridgeline galaxies 
brighter than $0.4L_{*}$ ($i$-band), within a scaled radius $r_{200}$
defined as the mean density of 200 times the mean density of the universe. 
SDSS spectroscopic redshifts are available for at least the brightest 
galaxy in $39\%$ of the clusters.  The sample is  up to $85\%$ complete 
for clusters with masses $\ge 10^{14}\cal{M}$$_{\odot}$.

\begin{figure}
   \centering
   \includegraphics[width=9cm]{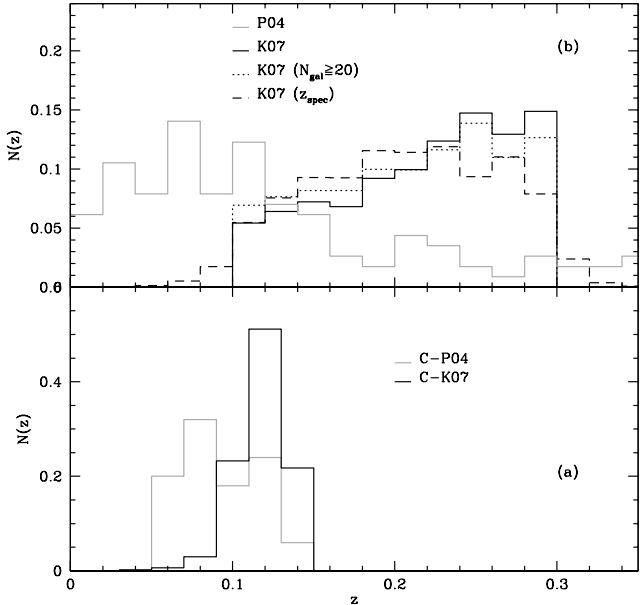}
      \caption{Panal (a) shows the spectroscopic redshiht distribution of the P04 galaxy clusters (grey solid line), the photometric redshift distribution of the K07 galaxy clusters (black solid line), the distributions of those K07 clusters with $N_{gal}\ge 20$ (dotted line) and the distribution of those K07 clusters with spectroscopic redshifts (dashed line). Panel (b) shows the redshift distributions of the selected clusters C-P04 (grey line) and C-K07 (black line).}
         \label{fig:cls_z}
\end{figure}

The sub-samples from P04 and K07 considered in this paper comprise galaxy clusters 
in the redshift range $0.05<z<0.14$. 
The upper limit was taken in order to have a volume limited sample of galaxies with an
adequate balance between the number of clusters, the range in absolute magnitude 
and the number of cluster members.
For K07 clusters, we applied a restriction 
in the richness, selecting clusters with $N_{gal}\ge 20$ in order to have cluster 
masses comparable to those in the P04 sample. The sub-samples are labeled as C-P04 and C-K07 
respectively. To select cluster members and estimate the physical 
properties of clusters, we use the Main Galaxy Sample (MGS, \citealt{Strauss:2002})
of the Fifth Data Release (DR5) of SDSS \citep{dr5} that is complete
down to a \citet{petro76} magnitude $r=17.77$.
At this point, our sub-samples comprise 54 X-ray galaxy 
clusters from the P04 sample and 612 from the K07 sample. We show in panel (a) of Figure \ref{fig:cls_z} the spectroscopic redshift distribution of 114 galaxy clusters of the P04 sample (grey solid line), the photometric redshift distribution of 13823 galaxy clusters of the K07 sample (black solid line), the distributions of those K07 clusters with $N_{gal}\ge 20$ (dotted line) and the distribution of those K07 clusters with spectroscopic redshifts (dashed line). Panel (b) shows the redshift distributions of the selected clusters C-P04 (grey line) and C-K07 (black line). We cross-correlated the catalogues to 
pick clusters that are included in both of them. If the projected 
distance between 2 clusters each belonging to a different catalogue 
is $<1{\rm Mpc}$, we consider that they are the same cluster. 
We found 20 coincidences; in these cases the clusters are only 
included in the X-ray sub-sample (C-P04).

In order to compute cluster properties such as velocity dispersion, we first
identified the cluster members. We use SDSS galaxies from the MGS
within a projected radius of $3{\rm Mpc}$ centred on the quoted cluster coordinates. 
From these galaxies we identify cluster members in two steps.
First, we use the friends-of-friends (\textit{fof}) algorithm developed by
\citet{H&G:1982} with the percolation linking length values according
to \citet{Diaz:2005}. As a result, we get for each field a list of 
substructures with at least 10 members identified by \textit{fof}.
The second step consists of eyeball examination of the structures
detected by \textit{fof}, a comparison between them and 
the listed cluster position and redshift to determine which 
coordinates and redshift fit better the observed galaxy over-density. 
From the redshift distribution of galaxies within
$|cz-cz_{\rm cluster}|\le 3000\kms$ we determine the line-of-sight 
extension of each cluster, i.e., a maximum and a minimum redshift for the
cluster. We then consider as cluster members all galaxies in the field that 
are within that redshift range. 
The \citet{Diaz:2005} technique includes an iterative method that provides 
precise dynamical centres of the clusters.  When these centres differ from
those computed by \citet{Koester:2007}, we adopted the \textit{fof} values.
We found that for $\sim 40\%$ for the clusters, the angular position given by
the \textit{fof} is different from the original value, whereas for 
$\sim 17\%$ of the clusters the redshift derived from the \textit{fof} algorithm
is a better match to the observed galaxy redshift distribution than those quoted 
in the original catalogues.
Finally, we excluded from our sample those clusters for which the \textit{fof}
redshift value lay outside the redshift range we consider in this paper.

\begin{figure}
   \centering
   \includegraphics[width=9.2cm]{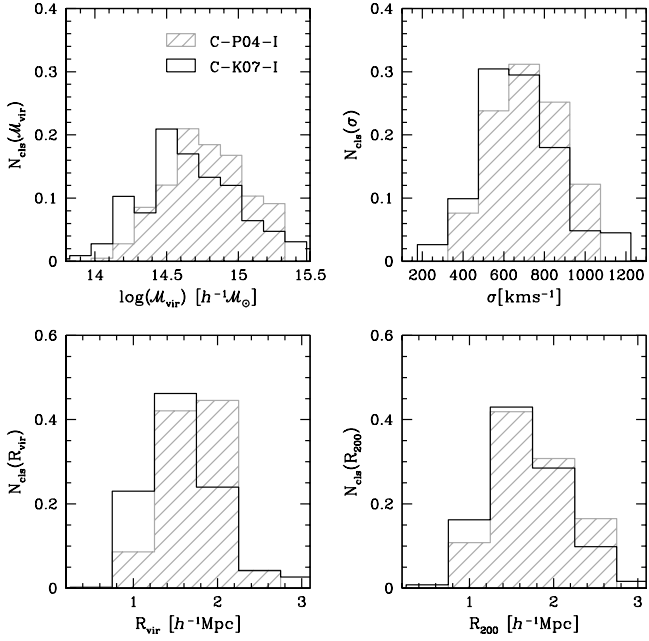}
      \caption{Distributions of cluster physical properties. The 
solid grey lines show the C-P04-I sample while the dashed histograms correspond to the 
C-K07-I sample.}
         \label{fig:hcls}
\end{figure}

Once the members of each cluster are selected, we compute cluster physical
properties we are interested in. We compute the line-of-sight
velocity dispersion $\sigma$, the virial radius and mass and the $r_{200}$. The line-of-sight velocity dispersion $\sigma$ is estimated using 
methods described by \citet{Beers:1990}. The bi-weight estimator was applied 
to clusters with $\ge 15$ members, whereas the gapper estimator was applied 
to poorer clusters. The radius $r_{200}$ was computed using the approximation 
provided by \citet{Carlberg:1997}. In a first step, 
we determined these parameters using all the cluster members within a $3{\rm Mpc}$ projected distance.
After this step, we recalculated $\sigma$ and $r_{200}$ using only 
galaxies located inside $r_{200}$. Virial radis and masses were 
computed following \citet{Merchan&Zandivarez:2005}.
The mean values of these parameters are shown in Table \ref{tb:mean},
where it can be seen that the sample C-P04 includes
on average clusters slightly more massive and larger than the sample
C-K07. Figure \ref{fig:hcls} shows the distributions of cluster physical 
properties of our samples. The distributions of the main properties of the 20 coincident clusters are similar to those of the parent samples.

\begin{table}
\center
\begin{tabular}{ccccc}
\hline \hline
  Sample         & $\sigma$ & $R_{200}$ & $\cal{M}$$_{vir}$        & $R_{vir}$ \\
           & [$\kms$] & [$\mpc$]  & [$h^{-1}\cal{M}$$_{\odot}$] &[$\mpc$]   \\
\hline
C-P04-I & $715$    & $1.77$    & $7\times10^{14}$      & $1.75$    \\
\hline
C-K07-I & $675$    & $1.67$    & $6\times10^{14}$       & $1.59$    \\
\hline \hline
\end{tabular}
\caption{Mean values of the cluster physical properties of our cluster samples.}
\label{tb:mean}
\end{table}

Through visual inspection we classified clusters based on their 
substructure. We only consider in our analysis sub-samples of regular 
clusters, that we label C-P04-I and C-K07-I, and we exclude systems that have two 
or more close substructures of similar size in the plane of the sky and/or in the 
redshift distribution. In order to compute the physical properties in a reasonable 
way, clusters with fewer than 10 galaxy members within $r_{200}$ are excluded 
\citep{Girardi:1993}. Our final galaxy cluster sample comprises 49 clusters 
from C-P04-I and 209 from C-K07-I.
Our sample of MaxBCG clusters is an order of magnitude smaller than the sample 
considered by \citet{Bernardi:2009}, also based on the \citet{Koester:2007} catalogue. This 
difference is mainly due to our restriction to select clusters with at least ten 
spectroscopic members in the redshift range  $0.05<z<0.14$. The sample of \citet{Bernardi:2009} includes clusters with redshift up to 0.3 and has, on average, fewer than 
one spectroscopic early-type satellite galaxy per cluster.

\subsection{The galaxy sample}
\label{subsec:galsample}

\begin{figure}
   \centering
   \includegraphics[width=8cm]{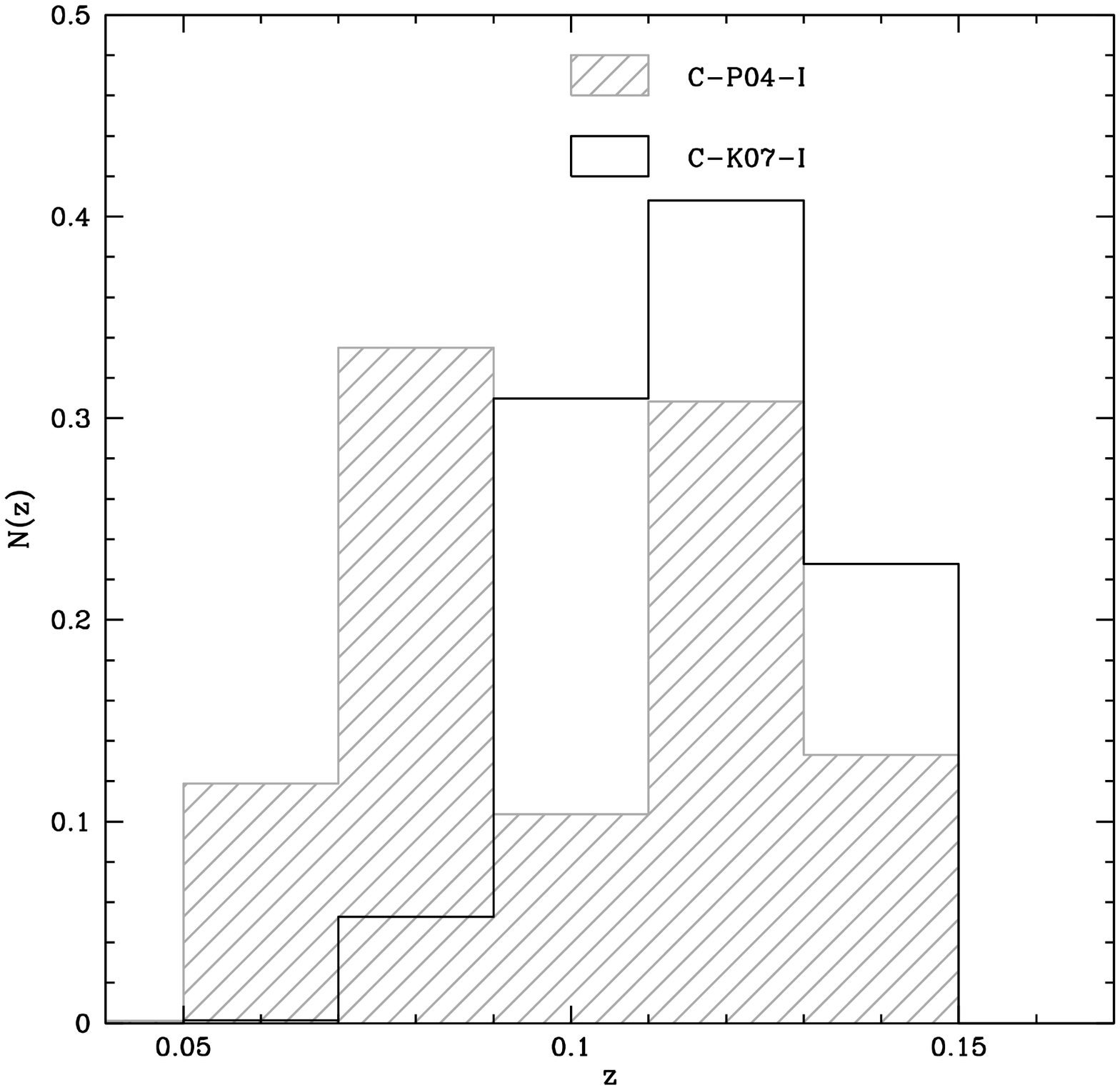}
      \caption{Redshift distributions of galaxies in our samples. Grey line: C-P04-I 
sample, black line: C-K07-I sample.}
         \label{fig:dz}
\end{figure}

In order to avoid the use of weights and since we are interested in
bright galaxies, we have constructed a volume limited samples instead of a flux limited 
sample. Taking into account the depth of our sample ($z<0.14$) and the apparent magnitude 
limit of the Main Galaxy Sample of the SDSS, our sample of galaxies only includes objects with  
$M_{^{0.1}r}<-21.3$. The C-P04-I sample comprises 786 galaxies, while the C-K07-I sample 
has 3041 galaxies. Figure \ref{fig:dz} shows the redshift distributions of galaxies of our samples, C-P04-I clusters with the grey line and C-K07-I sample with the black line.
Galaxy magnitudes used throughout this paper have been corrected for galactic extinction 
following \citet{sch98}, absolute magnitudes have been computed assuming 
$\Omega_0=0.3$, $\Omega_{\Lambda}=0.7$ and $H_0=70~h~{\rm km~s^{-1}~Mpc^{-1}}$ 
and $K-$corrected using the method of \citet{Blanton:2003}~({\small KCORRECT} 
version 4.1). All magnitudes are in the AB system.

Among the available data for each object in the MGS,
we have selected for our analyses parameters that are related to different 
physical properties of galaxies: luminosity, star formation rate, 
light distribution inside the galaxies and the dominant stellar populations.
In particular, the galaxy parameters we have focused our study on are the Petrosian $r-$band absolute 
magnitude $M_{^{0.1}r}$ and the radius that encloses 50\% of the Petrosian flux $r_{50}$.
We have chosen Petrosian quantities that are not corrected for the effects of seeing. In order to reduce the effect of the PS on galaxy sizes, we excluded galaxies with $r_{50}<2\arcsec$, i.e. $0.5\arcsec$ greater than the average seeing in SDSS. This is a conservative threshold, since the PSF in SDSS is known quite accurately \citep{Shen:2003}. \citet{HydeBernandi:2009} analysed residuals from the size-luminosity relation in the SDSS Petrosian $r$-band in different redshift ranges. As expected, they found that the effect of seeing increases as a function of redshift. However, in our redshift range ($0.05<z<0.14$), these residuals are close to $0$.

Our sample consists of bright galaxies in crowded fields where SDSS reductions tend to overestimate the sky level. This results in underestimates of magnitudes and half-light radius for large half-light radii galaxies (\citealt{dr6}, \citealt{Bernardi:2007}, \citealt{HydeBernandi:2009}). In order to correct the magnitudes and sizes by sky level, we have fitted curves to SDSS simulations (DR6 documentation) and we used these fits to correct the SDSS reductions. Briefly, SDSS quantified the sky effect by adding simulated galaxies (with exponential or de Vaucouleurs) profiles to SDSS images. The simulated galaxies ranged from apparent magnitude $r=12$ to $r=19$ in half-magnitude steps, with a one-to-one mapping from $r$ to S\'ersic half-light radius (\citealt{Sersic:1963,Sersic:1968}) determined using the mean observed relation between these quantities for MGS with exponential and de Vaucouleurs profiles. Axis ratios of 0.5 and 1 were used, with random position angle for the non-circular simulated galaxies. They found a difference between the input magnitude and the model magnitude returned by the SDSS photometric pipeline, as a function of magnitude. Their results are consistent with a separate analysis performed by \citet{HydeBernandi:2009}. Finally, we have estimated the rest-frame radius $r_{50}$ in $r$-band interpolating the observed radii in the adjacent $i$-band as in \citet{HydeBernandi:2009}.

\begin{figure}
   \centering
   \includegraphics[width=9.3cm]{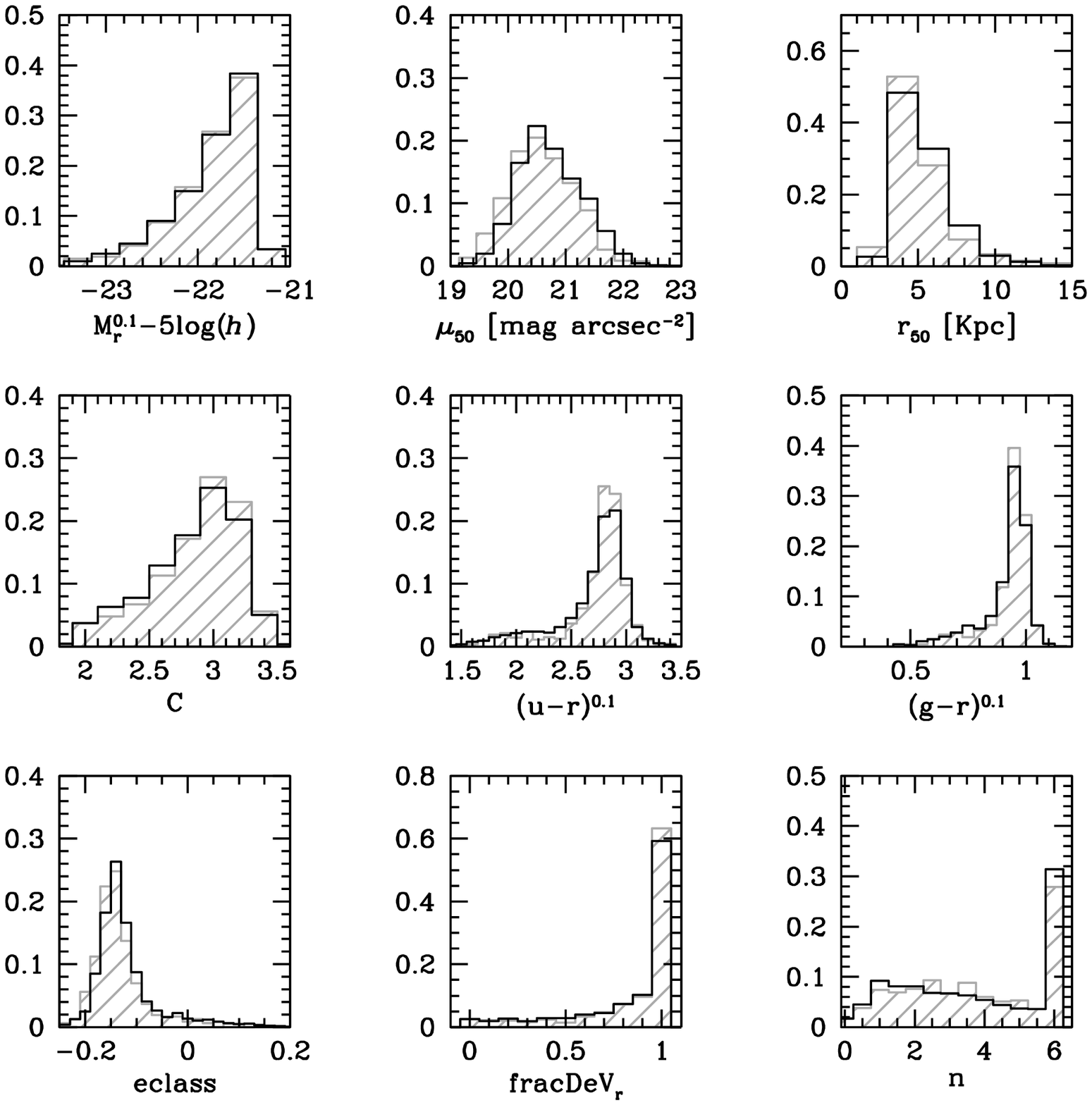}
      \caption{Distributions of galaxy properties of our samples. Grey line: C-P04-I sample, 
black line: C-K07-I sample.}
         \label{fig:par}
\end{figure}

Since the aim of this work is to analyze scaling relations and segregation of different 
types of galaxies in clusters and its comparison with field galaxies, we adopt several 
criteria to classify galaxies. \citet{Shima:2001} found that the concentration index shows a strong 
correlation with morphological type. These authors conclude that $C$ is perhaps the best 
parameter to classify the morphology of galaxies. They also found that galaxies with 
$R_{50}<2"$ show a weaker correlation between $C$ and the morphological type. To separate galaxies into early and late types, we use 
the $r-$band concentration index defined as the ratio
between the radii that enclose 90\%  and 50\% of the Petrosian flux,
$C=r_{90}/r_{50}$. Typically, early-type galaxies have $C>2.5$, while
for late-types $C<2.5$ \citep{Strateva:2001}. 
We also use the color to discriminate between early and late type galaxies. The corresponding 
threshold is $^{0.1}(g-r)=0.8-0.03(M_r +20)$ (see \citet{Blanton:2007}). 
Other indicator of galaxy type is the mono-parametric spectral classification based on the 
eigentemplate expansion of galaxy's spectrum ${\rm eclass}$. This parameter ranges from 
about $-0.35$ for early-type galaxies to $0.5$ for late-type galaxies (\citealt{Yip:2004}); the separation is set at $eclass=-0.1$. Finally, we use the S\'ersic index $n$, taken from \citet{Blanton:2005}. The distribution
of $n$ shows a bimodal distribution (see for instance \citet{Ball:2008} and \citet{Driver:2006}).
The $n$ separation is set at $2.5$, the value that divides in equal parts the two distributions and
corresponds to objects with approximately equal bulge and disk components. 
In figure \ref{fig:par} we show the distributions of galaxy parameters. We can see 
that clusters of the C-P04-I sample (on average, slightly more massive than C-K07-I) have a higher 
fraction of red galaxies than those taken from the C-K07-I sample (89\% and 85\% respectively).

We are also interested in the study of the brightest cluster galaxies and the comparison with 
early-type galaxies in both clusters and in the field. We identify the spectroscopic brightest 
cluster galaxy ($BCG_s$) in our sample. The $BCG_s$ in not necessarily the true BCG due to the 
following reasons: i) As a consequence of the finite fiber size of the SDSS spectrograph, any 
two fibers on the same plate need to be spaced at least $55\arcsec$ apart. In the case of a fiber 
collision, objects are selected at random. ii) The MSG is limited 
at the bright end by the fiber magnitude limits ($r=14.5$), to avoid saturation and excessive 
cross-talk in the spectrograph. These two effects together cause the spectroscopic sample to 
become incomplete.

\begin{figure*}
	\subfigure{\includegraphics[width=9.3cm]{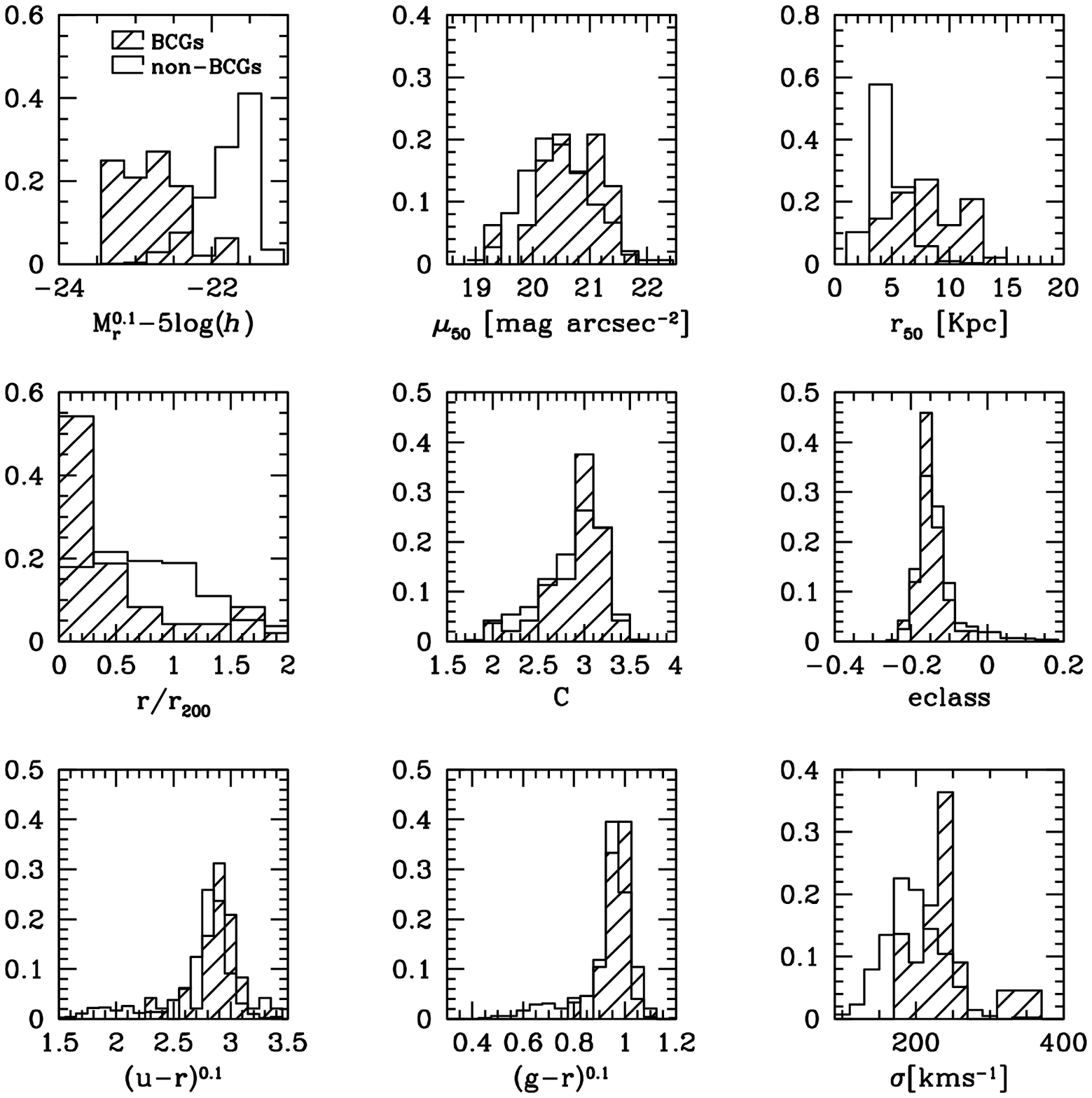}}
	\subfigure{\includegraphics[width=9.3cm]{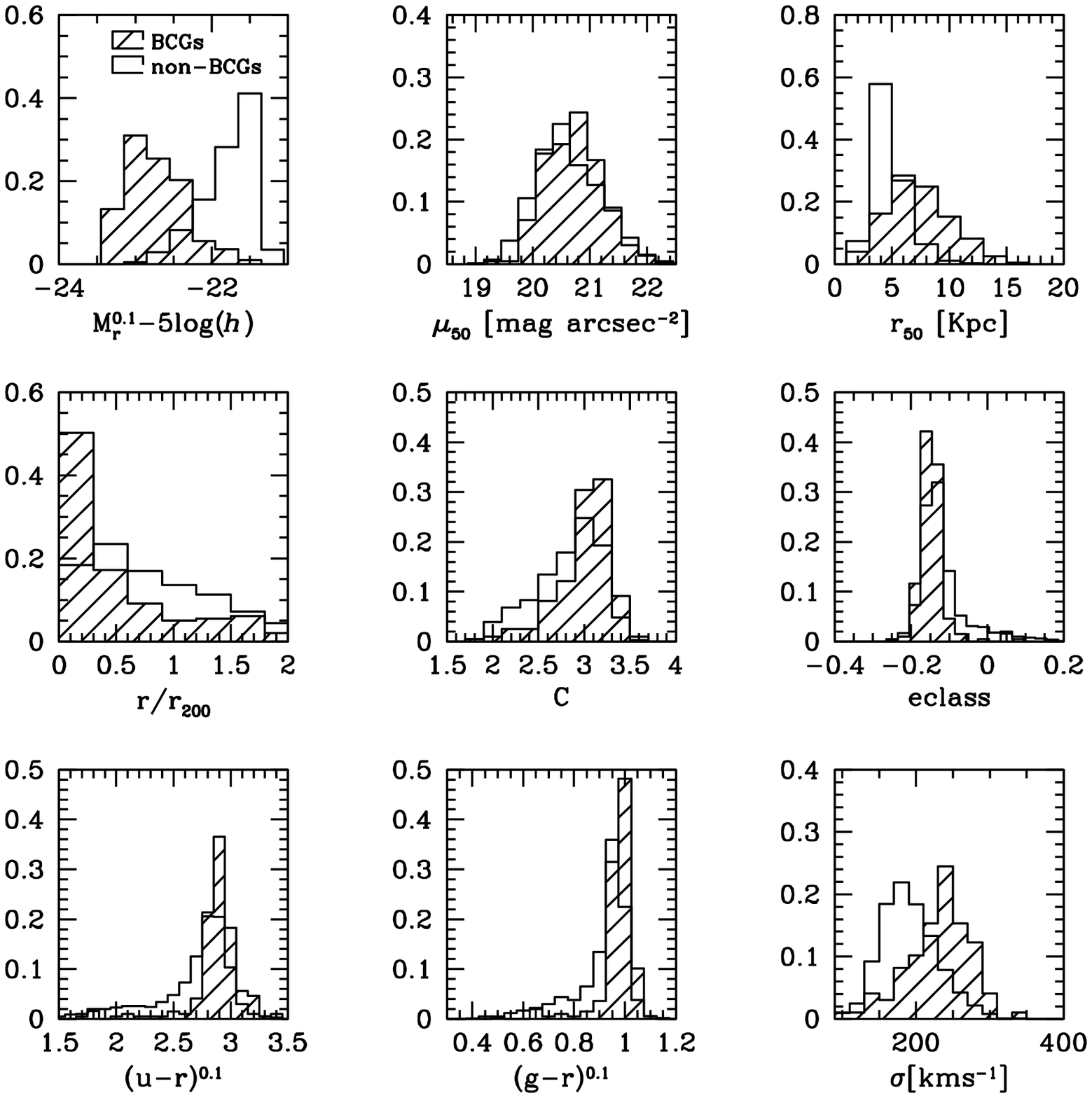}}\\
	\caption{Distributions of $BCG_s$ (dashed histograms) and $non-BCG_s$ galaxy properties of 
the samples C-P04-I (left panels) and C-K07-I (right panels).}
	\label{fig:bcg}
\end{figure*}

In figure \ref{fig:bcg} we show the distributions of $BCG_s$ parameters. We find that $BCG_s$ 
are larger and have lower surface brightness that non-$BCG_s$, in agreement with 
\citet{vonderlinden:2007}. In addition, $BCG_s$ have higher velocity dispersion than non-$BCG_s$. 
In general, we find that $BCG_s$ have colors, concentration and $eclass$ parameters typical of 
normal early-type galaxies. Despite that $BCG_s$ are usually in the centre of the cluster 
potential well, we find that the 25\% have $r/r_{200}>0.5$. Nevertheless, it is probable that a 
fraction of these $BCG_s$ with large $r/r_{200}$ are not the actual BCG. It is also 
possible that some of the cluster centres are wrong.
Finally, we do not find 
differences between the mean properties of $BCG_s$ in the C-P04-I and C-K07-I samples.

To perform a suitable comparisson between cluster and field, 
we construct a set of comparison samples. These were drawn from MGS and consist of galaxies
selected at random having the same redshift distribution as C-P04-I and C-K07-I.
This procedure excludes the cluster members.
For each cluster catalogue we have two types of samples: i) F-P04-T (52560 galaxies) and F-K07-T 
(51706 galaxies) that reproduce the corresponding total redshift distribution of galaxies in clusters 
and ii) F-P04-E, F-P04-L, F-K07-E and F-K07-L that reproduce the redshift distributions and 
concentration index distributions of the early and late type galaxy population of C-P04-I and 
C-K07-I, respectively.

\section{Scaling relations}
\label{sec:sr}


\subsection{Luminosity-size relation}
\label{ssec:lsr}

\begin{figure*}
   \centering
   \includegraphics[width=15cm]{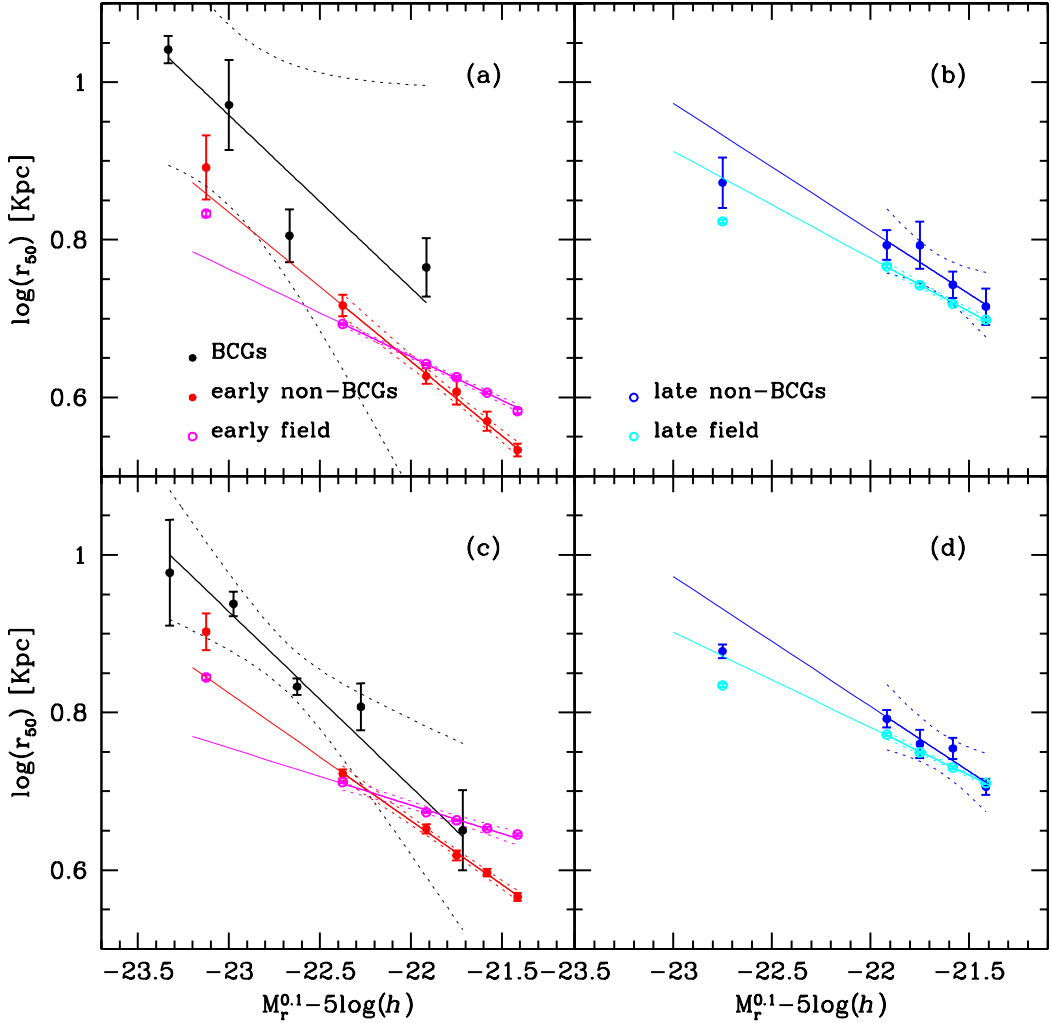}
      \caption{The luminosity size relation: Median of $\log(r_{50})$ versus $M_r^{0.1}$ 
for the C-P04-I sample (top panels) and the C-K07-I sample (bottom panels). Error bars were obtained 
by the bootstrap re-sampling technique. Dotted lines show the $95\%$ confidence bands of the linear 
regression. The red and magenta lines show the linear regression of non-$BCG_s$ early types in cluster 
and field respectively. The blue and cyan lines correspond to late types in cluster and field 
respectively. Black lines show the linear fits for $BCG_s$.}
         \label{fig:ls}
\end{figure*}

In this section we use the $r$-band concentration index $C$ to split galaxies into early and late types, 
as explained in section \ref{subsec:galsample}. In Figure \ref{fig:ls} we show the 
correlation between luminosity and the Petrosian half-light radius ($r_{50}\propto L^{\alpha}$). We plot the median values 
of $\log(r_{50})$ versus $M_r^{0.1}$ for the C-P04-I sample (top panels) and the 
C-K07-I sample (bottom panels). Error bars were obtained by the bootstrap re-sampling technique. 
Slopes and zero points in this work correspond to a least-square linear regression.
The dotted lines show the $95\%$ confidence bands of the linear regressions
. Galaxies in X-ray and MaxBCG selected clusters show similar size-luminosity relations.
Early and late type galaxies (red and blue lines respectively) show clear differences, late type galaxies being larger than early types for fixed luminosities. If clusters and field galaxies 
are compared, we can see in Figure \ref{fig:ls} and Table \ref{tb:lumsiz} that late type galaxies 
in the field (cyan lines) are smaller than galaxies in clusters for a fix luminosity.

For early type galaxies (magenta lines), the differences between field and cluster galaxies 
are also present, although, they are smaller. The effect is stronger for MaxBCG clusters. Finally, 
in the common range of luminosity, $BCG_s$ (black lines) tend to be larger and show a steeper 
$\log(r_{50})$ - $M_r^{0.1}$ relation than non-BCG early type galaxies in clusters. 
Except for $BCG_s$, early type galaxies in clusters and in the field show the previously 
reported curvature 
in the size-luminosity relation (\citealt{Bernardi:2007}). For these samples of galaxies, linear 
fits exclude the brightest luminosity bin.

We compare our results of the $r_{50}-M_r^{0.1}$ relation of BCGs with other studies. \citet{vonderlinden:2007} performed isophotal photometry from the C4 cluster catalogue (\citealt{Miller:2005}) and they found a slope $\alpha=0.65\pm 0.02$ for BCGs and $\alpha=0.63\pm 0.02$ for a control sample. \citet{Liu:2007} derived the luminosity-size relation for 85 BCGs from the same catalogue performing isophotal photometry down to four isophotal limits (22, 23, 24 and 25$\,mag\,arcsec^{-2}$); they found that the luminosity-size relation become steeper as the isophotal limit becomes fainter for both BCGs and control sample galaxies. The slope varies between $0.63\pm0.04$ and $0.88\pm0.04$ for BCGs and between $0.58\pm0.04$ and $0.74\pm0.04$ for the control sample. Although these authors used a similar photometry to \citet{vonderlinden:2007}, the difference in the slopes could be related to the differences in the brightness of the samples.
\citet{Bernardi:2007} fitted a de Vaucouleurs model to SDSS images from the C4 catalogue and used the half-light radius of their best-fitting models to study the luminosity-size relation. These authors found a slope $\alpha=0.89$. \citet{Lauer:2007} used surface photometry in HST images for 219 early type galaxies and  also fitted a de Vaucouleurs model; they found $\alpha=1.18\pm 0.06$. Once again, the photometry used in the last two works is similar but the results are very different. Again, a suitable explanation is the difference in the construction of the samples.
On the other hands, \citet{Guo:2009} used 911 central galaxies from SDSS DR4 Group catalogue (\citealt{Yang:2007}). This catalogue is constructed using the New York University Value-Addes Galaxy Catalog (\citealt{NYU}). \citet{Guo:2009} fitted a S\'ersic luminosity profile to each galaxy using GALFIT (\citealt{GALFIT}), and found a slope $\alpha=1.02\pm 0.03$ over the luminosity range $-19$ to $-24$. For galaxies brighter than $-22$, these authors found $\alpha=0.82\pm 0.06$. \citet{Shen:2003}, using spectroscopic SDSS galaxies and also S\'ersic luminosity profiles, found $\alpha=0.65$.
Recently, \citet{HydeBernandi:2009} used early-type BCGs identified in the SDSS MaxBCGS and C4 catalogs and found $\alpha=0.6$ for early-type galaxies, while BCGs follow a steeper relation.
It should be noted that the slopes resulting from our work are typically shallower than the slopes found in the different works mentioned above. Partially, these differences in slopes could be the result of the different methods used to perform the photometry. While the works mentioned above used de Vaucouleurs, isophotal or S\'ersic models, we used Petrosian magnitudes and sizes. In addition, the sample selection causes variations in the value of the slopes; for example the use or not of spectroscopic data can introduce important differences in the results, as has been demonstrated by \citet{Coenda:2006}. Beyond the differences in the slopes, it should be noted that BCGs always follow a steeper relation than galaxies in control samples.

\begin{table},
\begin{center}
\begin{tabular}{llll}
\hline \hline
          & C-P04-I         &                  &                  \\  
\hline
          & \it{early non-BCGs} & \it{BCGs}    & \it{early field}    \\
\hline
$\alpha$  & $0.48\pm 0.05$  & $0.55\pm 0.08$   & $0.425\pm 0.008$  \\
$b$       & $-3.5\pm0.3$    & $-4.1\pm0.6 $    & $-3.10\pm 0.07$  \\
\hline
          & \it{late}       &                  & \it{late field}         \\
\hline 
$\alpha$  & $0.4\pm 0.2$ &                     & $0.278\pm 0.005$     \\
$b$       & $-3\pm 1$       &                  & $-1.79\pm 0.04$    \\
\hline \hline
          & C-K07-I         &                  &                         \\ 
\hline
          & \it{early non-BCGs}     & \it{BCGs} &  \it{early field}       \\
\hline
$\alpha$  & $0.41\pm 0.02$    & $0.55\pm 0.08$ & $0.180\pm 0.003$       \\
$b$       & $-2.9\pm 0.2$     & $-4.2\pm 0.7$   & $-0.91\pm 0.03$     \\
\hline
          & \it{late}         &                 &   \it{late field}         \\
\hline
$\alpha$  & $0.40\pm 0.08$   &                  & $0.30\pm0.01$       \\
$b$       & $-2.8\pm 0.6$     &                 & $-1.88\pm0.08$      \\
\hline \hline
\end{tabular}
\end{center}
\caption{Parameter of the fits $\log(r_{50})= -0.4\alpha M_r+b$, see figure \ref{fig:ls}.} 
\label{tb:lumsiz}
\end{table}

\subsection{Dynamical relations}
\label{ssec:dyn}

\begin{figure*}
   \centering
   \includegraphics[width=13cm]{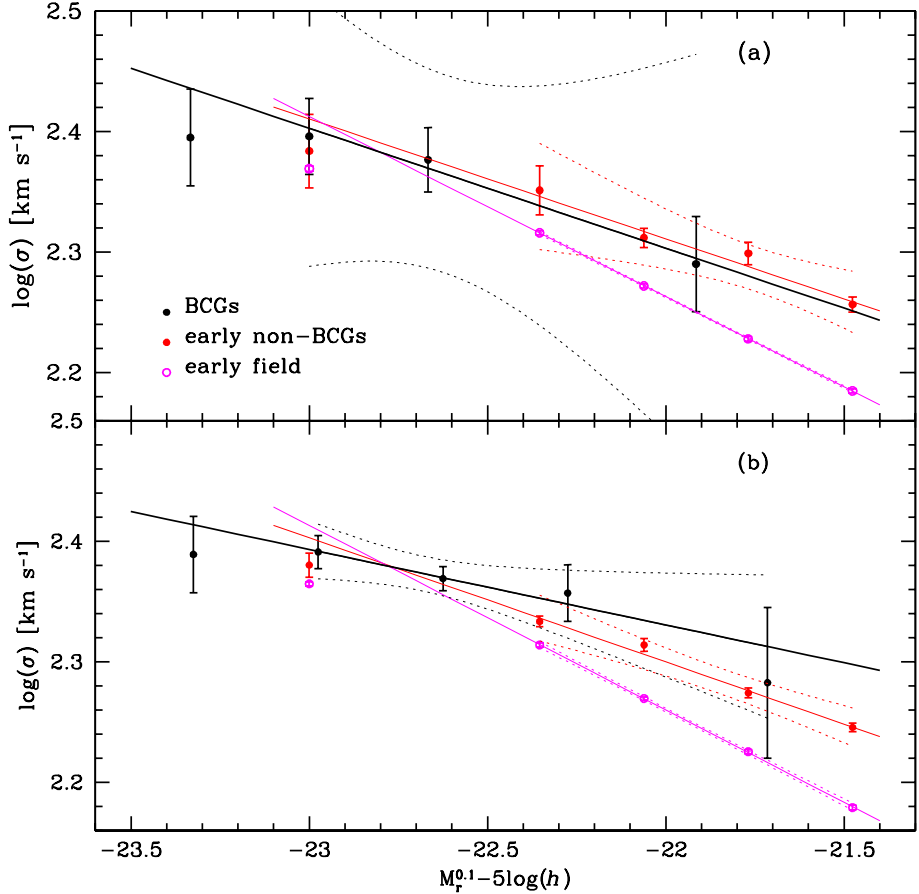}
      \caption{Faber-Jackson Relation: Linear fits of $\log(\sigma)$ versus $M_r^{0.1}$ for
$non-BCG_s$ early-types (red lines), $BCG_s$ (black lines) and control samples (magenta lines). The panels correspond 
to the same cases as in Figure \ref{fig:ls}.}
         \label{fig:fj}
\end{figure*}

It is known that velocity dispersion and luminosity are well correlated for early-type galaxies 
($L\propto \sigma^{\beta}$), a relation known as Faber-Jackson (\citealt{FJ:1976}). 
Figure \ref{fig:fj} shows the median values of $\log{\sigma}$ as function of $M^{0.1}_r$ 
for early-type galaxies in clusters (red lines), control sample (magenta lines) 
and $BCG_s$ (black lines). Table \ref{tb:fj} gives the parameters of the fits, where we can appreciate
that there are no differences between fits corresponding to both cluster samples. 
Where we do see a clear difference is between galaxies in clusters (both $BCG_s$ and 
non-$BCG_s$) and field galaxies, the latter being dynamically cooler (show lower mean velocity 
dispersions). The $\log{\sigma}$-$M^{0.1}_r$ also presents a departure from linearity 
for the brightest bin, where bright early type galaxies show a lower velocity dispersion than that
expected from the linear relation that fits the lower magnitude bins. $BCG_s$ show a $\log{\sigma}$-$M^{0.1}_r$ relation flatter than that of non-BCG galaxies. 
Nevertheless, the mean velocity dispersion that corresponds to the brightest luminosity bin of non-BCG 
galaxies is consistent with the values expected for $BCG_s$. Therefore, the flattening of the 
$\log{\sigma}$-$M^{0.1}_r$ relation observed for $BCG_s$ (see also \citet{Bernardi:2007}) 
could be a general property of early type galaxies regardless of its nature as the brightest cluster galaxy.
\citet{Liu:2007} and \citet{vonderlinden:2007} found that BCGs have a Faber-Jackson relation steeper than that of early-type galaxies. On the contrary, \citet{Bernardi:2007} found a shallower relation for BCGs, consistent with our findings.

\begin{figure*}
	\centering
        \includegraphics[width=13cm]{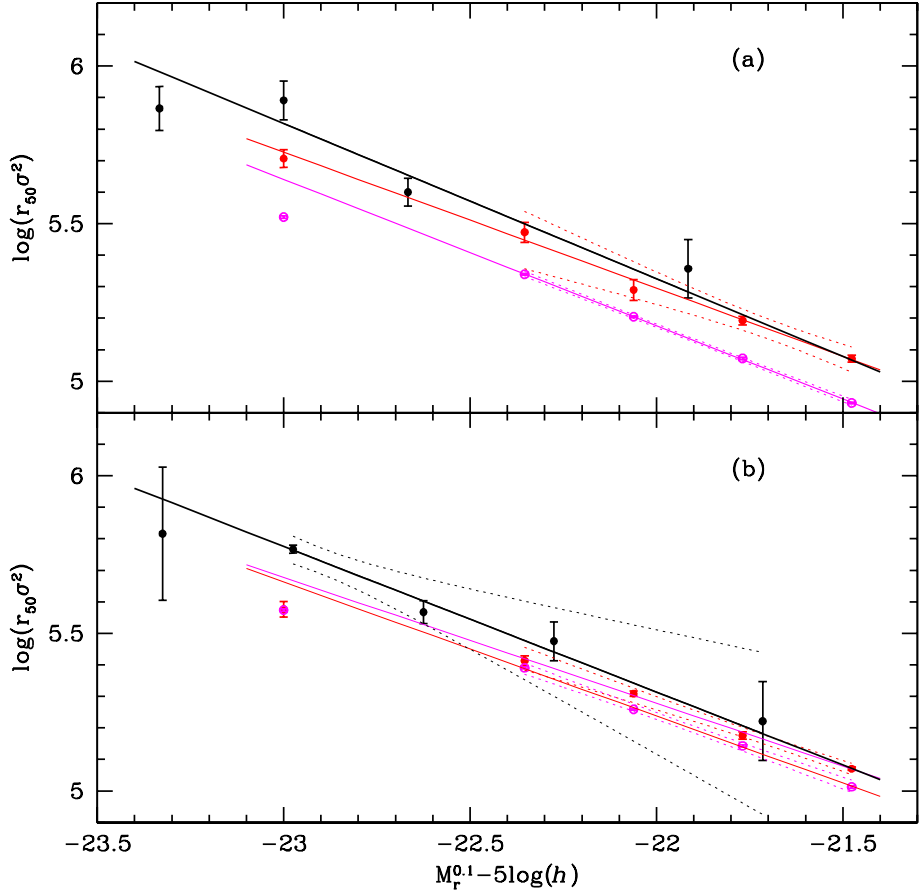}
	\caption{Correlation between dynamical mass $r_{50\sigma^2}$ and luminosity for
early-types (red lines), $BCG_s$ (black lines) and control samples (magenta lines), as in Figure 
\ref{fig:fj}.}
	\label{fig:ml}
\end{figure*}

In Figure \ref{fig:fj} we found that early-type and $BCG_s$ galaxies 
have, on average, the same velocity dispersions at a given luminosity. We also found that
$BCG_s$ have larger sizes than early type galaxies. Therefore, a different trend of dynamical 
masses as a function of luminosity between these two types of galaxies should be expected. In Figure \ref{fig:ml} we show the median value of $\log(r_{50}\sigma^2)$ as a function of the luminosity for early-types, $BGG_s$ and control-sample galaxies. Again, the brightest bins show a departure from linearity and were excluded from the linear fits. For the three samples we plot a linear regression that confirm
that $BCG_s$ have, on average, higher dynamical masses than early-type galaxies of the same magnitude
(the fitting parameters are given in Table \ref{tb:ml}). It should be noted that these differences are 
more significant for MaxBCG clusters (C-K07-I) than for X-ray clusters (C-P04-I). Field galaxies
in control samples show lower dynamical masses than non-BCGs in clusters.

\begin{table}
\begin{center}
\begin{tabular}{llll}
\hline \hline
          & \it{early non-BCGs} & \it{BCGs}      &    \it{early field}      \\
\hline 
          & C-P04-I           &                  & F-P04-E  \\  
\hline
$\beta$   & $4.0\pm 0.8$      & $4\pm 2$         &  $2.67\pm 0.05$  \\
$b$       & $-0.1\pm 0.3$     & $0\pm 1 $        &   $-1.03\pm 0.06$   \\
\hline \hline
          & C-K07-I           &                  & F-K07-E  \\ 
\hline
$\beta$   & $3.9\pm 0.2$      & $7\pm 3$         & $2.61\pm 0.05$ \\
$b$       & $0.0\pm 0.1$      & $1.0\pm 0.7$     & $-1.11\pm 0.06$    \\
\hline \hline
\end{tabular}
\end{center}
\caption{Parameters of the fits $\log(\sigma)= -\frac{0.4}{\beta}M_r+b$, see figure \ref{fig:fj}.} 
\label{tb:fj}
\end{table}

\begin{table}
\begin{center}
\begin{tabular}{llll}
\hline \hline
        & \it{early non-BCGs}     & \it{BCGs}     &   \it{early field}      \\
\hline 
        & C-P04-I        &                & F-P04-E  \\  
\hline
$a$   & $-0.43\pm 0.03$ & $-0.5\pm 0.1$    &   $-0.464\pm 0.005$  \\
$b$   & $-4.2\pm 0.7$   & $-6\pm 2 $       &   $-5.0\pm 0.1$   \\
\hline \hline 
      & C-K07-I        &                   & F-K07-E  \\ 
\hline
$a$   & $-0.40\pm 0.01$ &   $-0.46\pm 0.05$ & $-0.425\pm 0.005$ \\
$b$   & $-3.5\pm 0.3$     & $-5\pm 1$       & $-4.1\pm 0.1$    \\
\hline \hline
\end{tabular}
\end{center}
\caption{Parameters of the fits $\log(r_{50}\sigma^2)= aM_r+b$, see figure \ref{fig:ml}.}
\label{tb:ml}
\end{table}


\section{Galaxy segregation}
\label{sec:gs}

Using the C-P04-I and C-K07-I samples, \citet{MCM:2008} found that galaxy properties show 
a clear dependence on the clustocentric distance: a higher fraction of early type galaxies  
are found in the inner regions of clusters. In addition, they found that the $g-r$ colour is the 
property most predictive of the clustocentric distance of galaxies. \citet{Skibba:2008} used
SDSS galaxies with data from the Galaxy Zoo (\citealt{Bamford:2008}) and found that 
for fixed morphology, the environmental dependence of colour remains strong.    
\citet{Blanton:2007} found that the relation between galaxy colors and 
the distance from the centre of groups has a residual relationship on the clustering
of galaxies at small scales ($<300 h^{-1}$ Kpc).
In this work we are interested in the study of the segregation of several properties of bright 
galaxies in clusters as a function of $r/r_{200}$ and its dependence on the cluster 
identification techniques. \citet{Coenda:2006} found that these type of 
analysis can produce different results depending on whether redshift-confirmed members or 
galaxies in projection are considered. Although we are only considering bright galaxies, our 
analysis is based on confirmed members in volume limited galaxy samples identified with two different criteria and galaxies extracted from the same galaxy catalogue. 
The characteristics of these samples make them suitable for an unbiased analysis of the properties 
of galaxies and its dependence on the cluster environment.

As we explained in section \ref{subsec:galsample}, we have adopted several criteria to classify galaxies 
into morphological classes or types. We have used in our analysis the following parameters: 
the $g-r$ colour, the concentration index $C$, the $eclass$ parameter and the Sersic index $n$. 
Although they are not independent, they are dominated by different physical properties of galaxies.

\begin{figure}
   \centering
   \includegraphics[width=8.5cm]{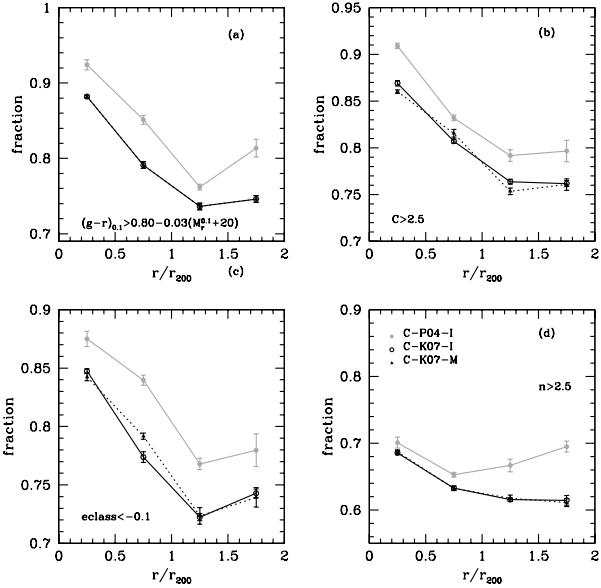}
      \caption{Fraction of early type galaxies as a function of the normalized clustocentric radius. 
Early type galaxies are selected according to several parameters: $(g-r)^{0.1}$ in panel (a), $C$ in 
panel (b), $eclass$ in panel (c) and $n$ in panel (d). The solid grey lines show the C-P04-I sample, 
whereas the solid black lines show the C-K07-I sample. The dotted black lines show the C-K07-M sample.}
         \label{fig:morseg}
\end{figure}

Figure \ref{fig:morseg} shows the fraction of early type galaxies as a function of the 
clustocentric radius. For both cluster samples a clear dependence of $r/r_{200}$ with $g-r$, $C$ 
and $eclass$ is found (panels a) to c)). The correlation is up to $r/r_{200} \sim 1$, for larger 
clustocentric distances we do not see any tendency. For the Sersic index $n$ 
(panel d)) 
we only observe a dependence for the C-K07-I sample. The comparison between the two cluster samples 
shows that the X-ray selected clusters always present a higher fraction of early type galaxies 
in the whole range of clustocentric distances.
As it was pointed out previously, the X-ray sample has a slightly higher fraction of massive clusters 
than the MaxBCG sample. In order to test if this difference in the mass distributions can be 
responsible for the higher fraction of early type galaxies in C-P04-I than in C-K07-I sample, we 
constructed a new sub-sample (C-K07-M) that consists of galaxies selected at random from C-K07-I 
having the same parent-cluster mass distribution as C-P04-I. As can be appreciated in figure 
\ref{fig:morseg}, the fractions of early type galaxies as a function of $r/r_{200}$ for C-K07-M 
(dotted lines) are indistinguishable from those corresponding to the C-K07-I sample, indicating 
that our finding of a higher fraction of early type galaxies in X-ray than in MaxBCG clusters is 
independent of the dynamical mass distribution.

\begin{figure}
   \centering
   \includegraphics[width=7.5cm]{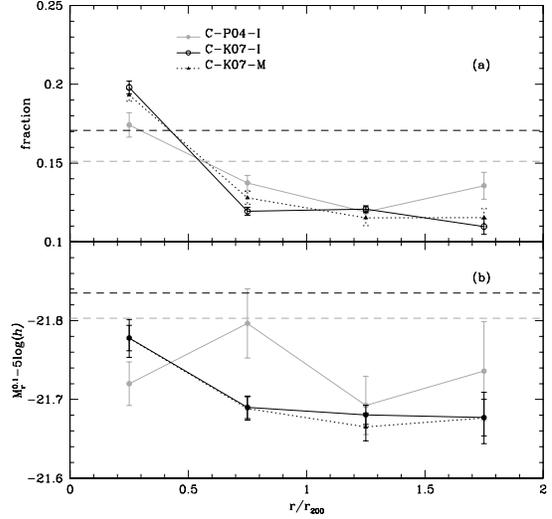}
      \caption{Panel (a): Fraction of galaxies brighter than $M_r^{0.1}\le -22.5$ as a function 
of $r/r_{200}$. The grey line shows the C-P04-I sample while the black line corresponds to the C-K07-I sample. The horizontal dashed lines show the median values of the corresponding field samples in the same colors as their cluster counterparts. Panel (b): Median values of $M_r^{0.1}$ versus $r/r_{200}$. The horizontal lines show the median values of the corresponding field samples.}
         \label{fig:fm}
\end{figure}

In panel (a) of figure \ref{fig:fm} we show the fraction of galaxies brighter than $M_r^{0.1}\le -22.5$ as a function of the normalized clustocentric distance. Analogously to the morphological segregation, we can see 
a clear dependence of this fraction for $r/r_{200} \lesssim 1$, in the sense that galaxies brighter $M_r^{0.1}\le -22.5$ 
preferentially inhabit the cluster centres. This trend is stronger for the C-K07-I sample. Alternatively, in panel (b) we show the median values of $M_r^{0.1}$ as a function $r/r_{200}$. We observe the dependence between the median $M_r^{0.1}$ and $r/r_{200}$ for the C-K07-I sample. This effect is not observed for C-P04-I sample. The dispersion observed in mean values of $M_r^{0.1}$ (especially for C-P04-I sample) is due to the natural spread of the luminosity function.

\begin{figure}
   \centering
   \includegraphics[width=7.5cm]{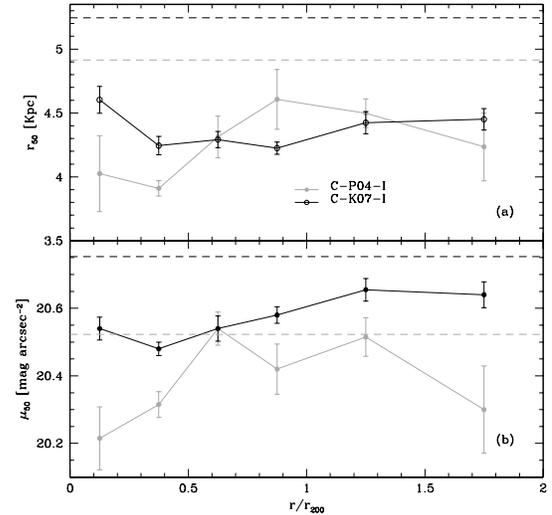}
      \caption{Panel (a) shows the median values of $r_{50}$ versus $r/r_{200}$ and panel(b) the median values of $\mu_{50}$ versus $r/r_{200}$. The horizontal lines show the median values of the corresponding field samples, as in Figure \ref{fig:fm}.}
         \label{fig:s0}
\end{figure}

\begin{figure}
   \centering
   \includegraphics[width=9.5cm]{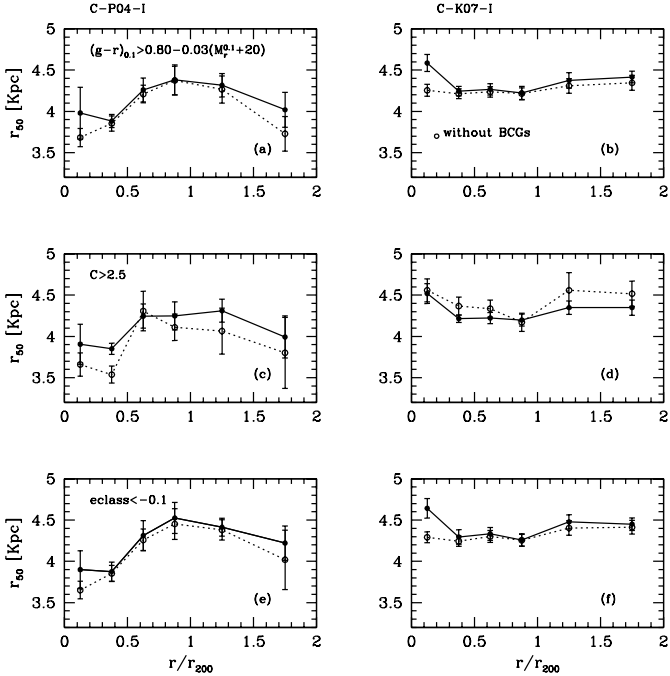}
      \caption{Median values of $r_{50}$ as a function of $r/r_{200}$ for early-type galaxies. In the 
left panels we show this relation for the C-P04-I sample while C-K07-I clusters are shown on the right 
panels. Early type galaxies were selected according to several parameters: $(g-r)^{0.1}$ in (a) and (b),  
$C$ in (c) and (d), and $eclass$ in (e) and (f). Dotted lines show the same relation but excluding 
the $BCG_s$.}
         \label{fig:s1}
\end{figure}

\begin{figure}
   \centering
   \includegraphics[width=9cm]{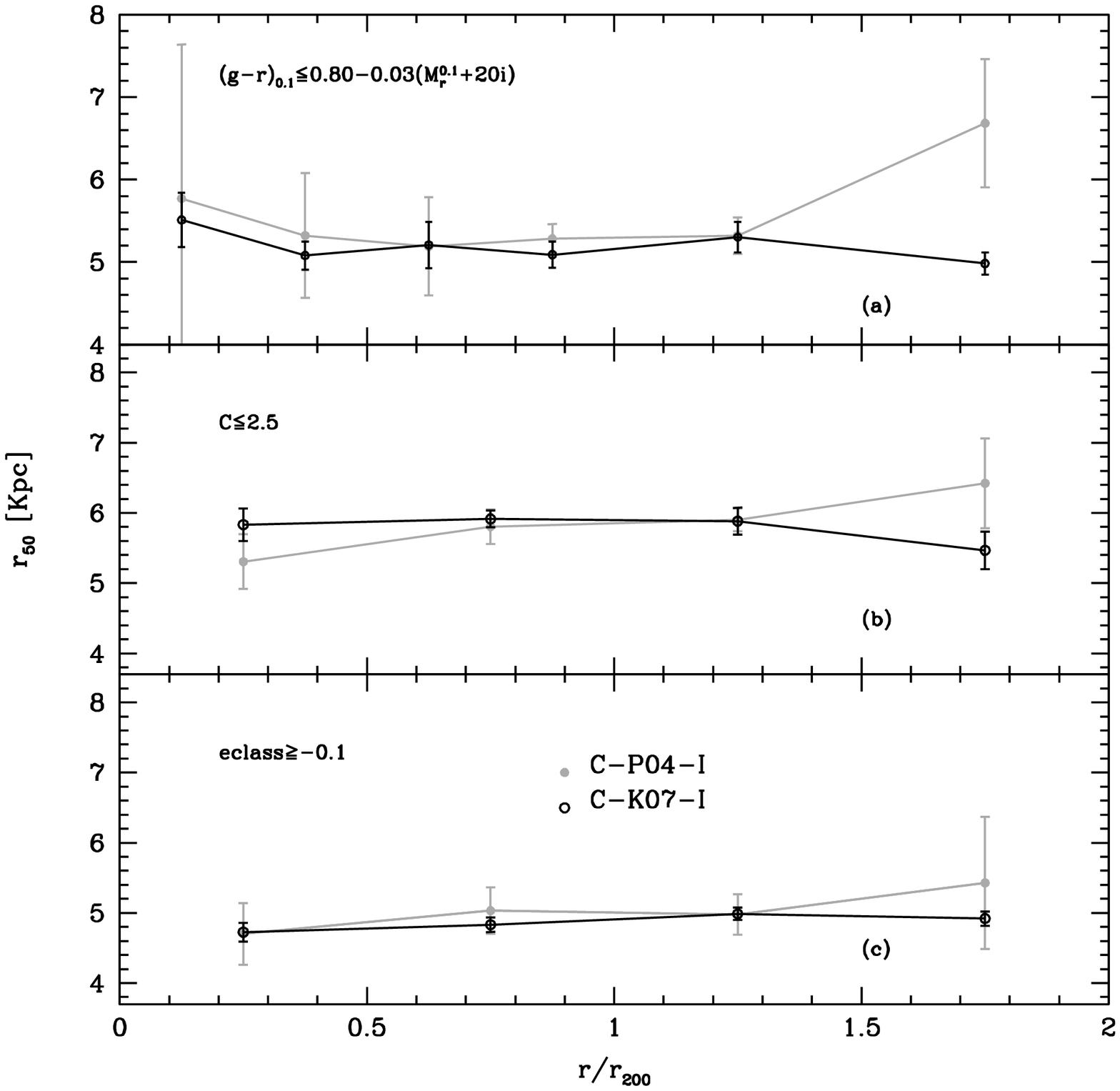}
      \caption{Median values of $r_{50}$ as a function of $r/r_{200}$ for late-type galaxies, 
analogously to figure \ref{fig:s1}.}
         \label{fig:s2}
\end{figure}

As discussed in section \ref{ssec:lsr}, the study of galaxy sizes and their relation to the 
luminosity helps to understand the galaxy formation and evolution. Here we study a possible 
dependence of galaxy size, $r_{50}$, as a function of $r/r_{200}$. Panel (a) in Figure \ref{fig:s0} shows the 
median value of $r_{50}$ versus the normalized clustocentric distance, for the C-P04-I sample 
(grey line) and for the C-K07-I sample (black line). 
For clustocentric distances larger than $0.8$, mean sizes of galaxies remain almost constant, being 
slightly larger for galaxies in X-ray clusters than in the MaxBCG sample, although the same effect 
is observed for field galaxies (horizontal dashed lines in figure \ref{fig:s0}).
For smaller clustocentric distances, median sizes show different behaviour: for galaxies in
the C-P04-I sample, median values of $r_{50}$ decrease as $r/r_{200}$ goes to zero, whereas the 
opposite behaviour is observed for galaxies in the C-K07-I sample. Panel (b) in Figure \ref{fig:s0} shows the median value of the $r-$band surface brightness $\mu_{50}$, computed inside $r_{50}$, as a function of $r/r_{200}$. For both cluster samples, $\mu_{50}$ decreases as $r/r_{200}$ goes to zero. Field galaxies have lower surface brightness (black horizontal line) than cluster members.

Figure \ref{fig:s1} repeats the previous analysis using only early type galaxies. In the 
left panels we show this relation for the C-P04-I sample while C-K07-I sample are shown in the right 
panels. We have also evaluated the influence of $BCG_s$ in our 
results, repeating the analysis without $BCG_s$ (dotted lines).
Analogously to Figure \ref{fig:s0}, the sizes of early type galaxies increase when $r/r_{200}$ 
decreases in the C-K07-I sample. Nevertheless, this tendency disappears when $BCG_s$ are excluded. 
On the other hand, the median values of $r_{50}$ decrease as $r/r_{200}$ goes to zero for the C-P04-I 
sample; this effect is stronger when $BCG_s$ are excluded. These results clearly show an important 
difference between clusters identified with different criteria. As can be appreciated in figure 3, 
a higher fraction of early type galaxies is observed for the X-ray selected clusters. Consequently,
the different behaviour observed in Figure \ref{fig:s1}, could be the result of the differences 
in the population of early type galaxies between the two samples. For instance, a higher fraction 
of elliptical galaxies in the early type population of the X-ray selected clusters with respect to 
the MaxBCG sample could produce a different behaviour in the mean sizes as a function of $r/ r_{200}$. 
In order to test this possibility, we repeated the analysis of Figure \ref{fig:s1} selecting early type 
galaxies with $c>3.1$. This new threshold includes a third of the early type galaxies previously 
analyzed, and it will mostly select elliptical galaxies. For this sub-sample of early type galaxies, 
we find the same trend observed in Figure \ref{fig:s1}, confirming that there is a dependence of the 
mean sizes of early type galaxies on the cluster selection criterion. Based on a 
sample of 228 elliptical galaxies in 24 clusters, \citet{Cypriano:2006} found that elliptical galaxies
in the inner regions of clusters are typically 5\% smaller than those in the outer regions. These authors
interpreted these results in terms of the tidal stripping of stars that lead to changes in the whole 
structure of galaxies in the central regions of clusters.

Figure \ref{fig:s2} is the analogous of Figure \ref{fig:s1} for late-type galaxies. Although galaxies 
in X-ray selected clusters show a tendency on average $r_{50}$ to be smaller towards the cluster
centers, this behaviour is not statistically significant.

\section{Discussion and conclusions}
\label{sec:conc}

We have compiled two catalogues of massive clusters of galaxies based on samples 
that make use of different identification techniques: the P04 sample that comprises bright 
X-ray galaxy clusters and the K07 sample, constructed 
according to the properties of early-type galaxies. We selected sub-samples of galaxy clusters in 
the redshift range $0.05<z<0.14$. For both samples, member galaxies are brighter that 
$M_r^{0.1}=-21.3$ and were identified from a spectroscopic volume-limited sample of 
the DR5-SDSS. In order to select a fair sample of clusters of galaxies we use the \textit{fof} 
algorithm and an eyeball examination of each overdensity identified by \textit{fof}. 
Throughout the visual inspection, we classified the clusters based on their substructure. 
We only consider in our analysis regular clusters and we exclude systems that have two 
or more close substructures of similar size in the plane of the sky and/or in the redshift 
distribution. In addition, we have determined new positions in the plane of the sky and in 
the line of sight. We found that for $\sim 40\%$ of the clusters, the angular position given 
by \textit{fof} is better than the original value, whereas for $\sim 17\%$ of the clusters, 
the redshift given by \textit{fof} is a better match to the observed mean redshift distribution 
than the original value. Our final sample comprises $49$ X-ray selected clusters with 786 
galaxies and $209$ MaxBCG clusters with 3041 galaxies. We have determined the following 
physical properties of clusters: the line-of-sight velocity dispersion $\sigma$, the virial 
radius and mass and $r_{200}$.

We have analyzed different scaling relations between photometric and dynamical parameters. 
Bright galaxies were separated in early and late types according to several criteria. We have also
studed the segregation of galaxies analysing different properties as a function of the normalized 
clustocentric distances. The main results are:
\begin{itemize}
\item Galaxies in X-ray and MaxBCG selected clusters show similar size-luminosity relations.
\item We found that each galaxy type has a different luminosity-size relation. This 
implies a different mass-luminosity relation and/or a different star formation history.
\item For each luminosity bin, late type galaxies in the field have sizes smaller than their cluster counterpart.
\item The same effect is observed for early type galaxies, although the differences are smaller. 
The stronger effect corresponds to galaxies in MaxBCG clusters.
\item At fixed luminosity, $BCG_s$ tend to be larger and show a steeper $\log(r_{50})$ - 
$M_r^{0.1}$ relation than non-BCG early type galaxies in clusters.
\item Non-BCGs early type galaxies in clusters and in the field show the known curvature in the 
size-luminosity relation.
\item The Faber-Jackson relation for early-type galaxies in clusters is the
same in X-ray and MaxBCG clusters.
\item We found a clear difference between galaxies in clusters (both $BCG_s$ and 
non-BCG) and in the field, showing the later lower values of the velocity dispersion at fixed luminosity.
\item The $\log{\sigma}$-$M^{0.1}_r$ relation presents, at the bright end, a departure from 
linearity. Bright early type galaxies show a lower velocity dispersion than the
expected from the linear relation at lower magnitudes.
\item  $BCG_s$ show a flatter $\log{\sigma}$-$M^{0.1}_r$ relation than non-BCG galaxies. 
Nevertheless, the velocity dispersion of the brightest non-BCG galaxies are consistent with
the values expected for $BCG_s$.
\item At fixed luminosity, $BCG_s$ have, on average, higher dynamical masses than early-type 
galaxies in clusters.
The difference is more significant for MaxBCG clusters than for X-ray selected clusters.
\item At fixed luminosity, field early type galaxies in the control samples show lower dynamical 
masses than non-BCG 
early type galaxies in clusters.
\item Using several criteria to classify galaxies into morphological classes, we found the well 
know morphological segregation for both samples of clusters. The correlation is up to 
$r/r_{200} \sim 1$; for larger clustocentric distances we do not see any trend.
\item For the whole range of clustocentric distances, X-ray selected clusters have a higher 
fraction of early type galaxies than MaxBCG clusters.
\item  Within $r/r_{200} \sim 1$ we found a dependence of the fraction of galaxies brighter than 
$M_r^{0.1}\le -22.5$ as a function of the normalized clustocentric distance. Bright galaxies preferentially 
inhabit the cluster centers. This tendency is stronger for MaxBCG clusters.
\item For clustocentric distances larger than $0.8$, mean sizes of galaxies remain almost constant, 
being slightly larger for galaxies in X-ray selected clusters than in the MaxBCG sample.
\item For $r/r_{200} \leq 0.8$, median sizes depend on the cluster sample. 
For galaxies in X-ray selected clusters, median values of $r_{50}$ decrease as $r/r_{200}$ goes 
to zero, whereas the opposite is observed for galaxies in the MaxBCG clusters.
\item For MaxBCG clusters, the sizes of early type galaxies increase as $r/r_{200}$ decreases. The 
trend disappears when $BCG_s$ are excluded. On the contrary, the median value of $r_{50}$ 
decrease as $r/r_{200}$ goes to zero for X-ray selected clusters. The effect is stronger when 
$BCG_s$ are excluded. 
\end{itemize}

Early type galaxies in the field are lower, have lower velocity dispersion and have smaller virial 
masses than their cluster counterpart. On the other hand, $BCG_s$ are larger and have higher values of the virial 
masses than non-BCG in clusters, although the velocity dispersions are comparable. These differences 
in the properties of early type galaxies in the field, in clusters and with $BCG_s$ are consistent 
with the scenario where the environment plays a fundamental rol in the formation histories of 
galaxies. These results suggest three different formation scenarios, $BCG_s$ being the 
galaxies less affected by dissipation. Dry mergers have no energy loss mechanism, therefore the 
formed galaxies are less centrally contracted with optical sizes larger than those of galaxies that 
have undergone a dissipative merger (\citealt{Kormendy:1989}, \citealt{Nipoti:2003}, 
\citealt{Bernardi:2007}). \citet{Weinmann:2009} and \citet{Guo:2009} analyzed the size-luminosity relation of galaxies in groups, where the environmental effects are supposed to be different to those expected in clusters. These authors found the same size-luminosity relation for both central and satellite early type galaxies. The different results between galaxies in groups and clusters also support the idea that the size-luminosity relation depends on the environment.

At the same luminosity, late-type galaxies in clusters have larger sizes, and therefore lower 
surface brightness than their field counterparts (control sample). This is in 
agreement with the scenario of strangulation \citep{Larson:1980}, where the gas in the halo 
is stripped, suppressing the supply of cold gas and therefore affecting the star formation rate and 
the galaxy luminosity.

The X-ray selected sample C-P04-I is made of bright X-ray galaxy clusters, whereas the C-K07-I sample was 
constructed according to the properties of early-type galaxies.
Several clusters in the C-K07-I sample should be X-ray emitters, although they would be, 
on average, less X-ray luminous than C-P04-I clusters. The fact that X-ray selected 
clusters show a higher fraction of early type galaxies whose sizes tend to decrease towards the cluster 
center, effect that is not observed for galaxies in MaxBCG clusters, can be interpreted as possible 
evidence that the hot intracluster medium is also playing a role in the evolution of early type galaxies. 
Alternatively, the observed effect could be caused by differences in the clusters masses. Even though the 
effects mentioned above are also present for sub-samples of C-P04-I and C-K07-I with the same virial 
mass distribution, due to the usual uncertainties in the computation of these masses, we cannot discard 
differences in the actual mass distributions. If this is the case, and C-P04-I clusters are on average more 
massive than C-K07-I, decreases in the mean sizes of early type galaxies in X-ray selected clusters 
could be the result of tidal stripping that can produce changes in the structure of galaxies 
(see the results from numerical simulations of \citet{Aguilar:1986} and 
observational evidences from \citet{Trujillo:2002} and \citet{Cypriano:2006}).
This difference between X-ray and MaxBCG selected clusters is an unexpected result that 
should be explored in more detail.

\begin{acknowledgements}
This work has been partially supported with grants from Consejo Nacional
de Investigaciones Cient\'\i ficas y T\'ecnicas de la Rep\'ublica Argentina
(CONICET) and Secretar\'\i a de Ciencia y Tecnolog\'\i a de la Universidad 
de C\'ordoba. We kindly thank Dr. H\'ector J. Mart\'inez for his helpful comments on the manuscript, which contributed to improve the present paper. VC acknowledges the guidance of Dr. Ariel Zandivarez in 
the construction of the cluster sample.
\end{acknowledgements}

\bibliographystyle{aa} 
\bibliography{coenda} 


\Online

\begin{appendix}
\section{Tables}

\tiny
\longtab{1}{
\begin{longtable}{llccccc}
\caption{\label{tb:pop} C-P04-I cluster sample.}\\
\hline\hline
$\alpha$ (J$2000$.$00$)& $\delta$ (J$2000$.$00$) & $z$ & $\sigma$ &$M_{vir}$  & $R_{vir}$ & $R_{200}$\\
 $[h\; m \; s]$ & [$\degr\; \arcmin\;\arcsec$]  &   & $[\kms$] & [$h^{-1}\Msol$]  & [$\mpc$] & [$\mpc$] \\
\hline
\endfirsthead
\caption{continued.}\\
\hline\hline
$\alpha$ (J$2000$.$00$)& $\delta$ (J$2000$.$00$) & $z$ & $\sigma$ &$M_{vir}$  & $R_{vir}$ & $R_{200}$ \\
$\alpha$ (J$2000$.$00$)& $\delta$ (J$2000$.$00$) & $z$ & $\sigma$ &$M_{vir}$  & $R_{vir}$ & $R_{200}$\\
 $[h\; m \; s]$ & [$\degr\; \arcmin\;\arcsec$]  &   & $[\kms$] & [$h^{-1}\Msol$]  & [$\mpc$] & [$\mpc$] \\
\hline
\endhead
\hline
\endfoot
$ 0\,\, 41\,\, 50.1  $ &$ - 9\,\, 18\,\,  6.8 $ & $ 0.056 $ & $  698.82 $ & $  0.558x10^{15} $ & $ 1.638 $ & $ 1.729 $ \\
$  1\,\, 19\,\, 37.7  $ &$ +14\,\, 53\,\, 35.2 $ & $ 0.129 $ & $  778.90 $ & $  0.502x10^{15} $ & $ 1.185 $ & $ 1.927 $ \\
$  7\,\, 36\,\, 25.0  $ &$ +39\,\, 25\,\, 58.4 $ & $ 0.117 $ & $  605.19 $ & $  0.482x10^{15} $ & $ 1.887 $ & $ 1.497 $ \\
$  7\,\, 53\,\, 19.0  $ &$ +29\,\, 22\,\, 26.8 $ & $ 0.061 $ & $  689.22 $ & $  0.606x10^{15} $ & $ 1.829 $ & $ 1.705 $ \\
$  8\,\,  9\,\, 40.2  $ &$ +34\,\, 55\,\, 34.3 $ & $ 0.082 $ & $  528.46 $ & $  0.323x10^{15} $ & $ 1.656 $ & $ 1.308 $ \\
$  8\,\, 10\,\, 22.6  $ &$ +42\,\, 16\,\,  0.8 $ & $ 0.064 $ & $  548.75 $ & $  0.292x10^{15} $ & $ 1.391 $ & $ 1.358 $ \\
$  8\,\, 22\,\, 10.0  $ &$ +47\,\,  5\,\, 58.2 $ & $ 0.130 $ & $  674.55 $ & $  0.671x10^{15} $ & $ 2.114 $ & $ 1.669 $ \\
$  8\,\, 25\,\, 27.6  $ &$ +47\,\,  7\,\, 10.6 $ & $ 0.126 $ & $  792.06 $ & $  0.952x10^{15} $ & $ 2.175 $ & $ 1.960 $ \\
$  8\,\, 28\,\,  6.7  $ &$ +44\,\, 45\,\, 48.2 $ & $ 0.145 $ & $  577.57 $ & $  0.187x10^{15} $ & $ 0.803 $ & $ 1.429 $ \\
$  8\,\, 45\,\, 29.0  $ &$ +44\,\, 34\,\, 28.2 $ & $ 0.054 $ & $  331.69 $ & $  0.545x10^{14} $ & $ 0.710 $ & $ 0.821 $ \\
$  9\,\, 13\,\,  9.2  $ &$ +47\,\, 43\,\, 26.4 $ & $ 0.052 $ & $  408.98 $ & $  0.120x10^{15} $ & $ 1.032 $ & $ 1.012 $ \\
$  9\,\, 53\,\, 41.5  $ &$ + 1\,\, 42\,\, 42.5 $ & $ 0.098 $ & $  403.63 $ & $  0.165x10^{15} $ & $ 1.454 $ & $ 0.999 $ \\
$ 10\,\, 23\,\, 41.1  $ &$ +49\,\,  8\,\,  5.6 $ & $ 0.144 $ & $  618.10 $ & $  0.532x10^{15} $ & $ 1.997 $ & $ 1.529 $ \\
$ 10\,\, 54\,\,  5.5  $ &$ +54\,\, 50\,\, 50.6 $ & $ 0.072 $ & $  448.67 $ & $  0.192x10^{15} $ & $ 1.368 $ & $ 1.110 $ \\
$ 10\,\, 58\,\, 26.3  $ &$ +56\,\, 47\,\, 31.9 $ & $ 0.136 $ & $  597.87 $ & $  0.332x10^{15} $ & $ 1.330 $ & $ 1.479 $ \\
$ 11\,\, 13\,\, 22.7  $ &$ + 2\,\, 32\,\, 32.6 $ & $ 0.075 $ & $  431.97 $ & $  0.170x10^{15} $ & $ 1.304 $ & $ 1.069 $ \\
$ 11\,\, 15\,\, 32.2  $ &$ +54\,\, 26\,\,  5.6 $ & $ 0.070 $ & $  628.67 $ & $  0.334x10^{15} $ & $ 1.213 $ & $ 1.556 $ \\
$ 11\,\, 21\,\, 36.2  $ &$ +48\,\,  3\,\, 50.0 $ & $ 0.112 $ & $  728.02 $ & $  0.541x10^{15} $ & $ 1.462 $ & $ 1.801 $ \\
$ 11\,\, 33\,\, 17.3  $ &$ +66\,\, 22\,\, 45.5 $ & $ 0.115 $ & $  847.57 $ & $  0.934x10^{15} $ & $ 1.865 $ & $ 2.097 $ \\
$ 11\,\, 44\,\,  8.2  $ &$ + 5\,\, 45\,\, 22.7 $ & $ 0.103 $ & $  586.83 $ & $  0.370x10^{15} $ & $ 1.541 $ & $ 1.452 $ \\
$ 11\,\, 44\,\, 40.8  $ &$ +67\,\, 24\,\, 40.0 $ & $ 0.117 $ & $  453.55 $ & $  0.221x10^{15} $ & $ 1.542 $ & $ 1.122 $ \\
$ 12\,\,  0\,\, 24.5  $ &$ + 3\,\, 19\,\, 51.6 $ & $ 0.133 $ & $  858.09 $ & $  0.119x10^{16} $ & $ 2.310 $ & $ 2.123 $ \\
$ 12\,\, 17\,\, 40.8  $ &$ + 3\,\, 39\,\, 41.0 $ & $ 0.077 $ & $  922.14 $ & $  0.113x10^{16} $ & $ 1.903 $ & $ 2.282 $ \\
$ 12\,\, 58\,\, 41.1  $ &$ - 1\,\, 45\,\, 24.8 $ & $ 0.084 $ & $  740.78 $ & $  0.769x10^{15} $ & $ 2.010 $ & $ 1.833 $ \\
$ 13\,\,  2\,\, 50.7  $ &$ - 2\,\, 30\,\, 22.3 $ & $ 0.083 $ & $  681.06 $ & $  0.534x10^{15} $ & $ 1.650 $ & $ 1.685 $ \\
$ 13\,\,  3\,\, 56.5  $ &$ +67\,\, 31\,\,  3.7 $ & $ 0.106 $ & $  659.69 $ & $  0.358x10^{15} $ & $ 1.179 $ & $ 1.632 $ \\
$ 13\,\,  9\,\, 17.0  $ &$ - 1\,\, 36\,\, 45.4 $ & $ 0.083 $ & $  519.51 $ & $  0.238x10^{15} $ & $ 1.264 $ & $ 1.285 $ \\
$ 13\,\, 26\,\, 17.8  $ &$ + 0\,\, 13\,\, 32.5 $ & $ 0.082 $ & $  501.04 $ & $  0.242x10^{15} $ & $ 1.382 $ & $ 1.240 $ \\
$ 13\,\, 30\,\, 49.9  $ &$ - 1\,\, 52\,\, 22.1 $ & $ 0.087 $ & $  659.17 $ & $  0.480x10^{15} $ & $ 1.585 $ & $ 1.631 $ \\
$ 13\,\, 32\,\, 38.9  $ &$ +54\,\, 19\,\,  9.5 $ & $ 0.101 $ & $  811.68 $ & $  0.927x10^{15} $ & $ 2.018 $ & $ 2.008 $ \\
$ 13\,\, 36\,\,  6.5  $ &$ +59\,\, 12\,\, 26.6 $ & $ 0.071 $ & $  934.40 $ & $  0.123x10^{16} $ & $ 2.017 $ & $ 2.312 $ \\
$ 13\,\, 42\,\, 28.3  $ &$ + 2\,\, 14\,\, 45.2 $ & $ 0.077 $ & $  752.26 $ & $  0.686x10^{15} $ & $ 1.738 $ & $ 1.861 $ \\
$ 13\,\, 53\,\,  0.8  $ &$ + 5\,\,  9\,\, 21.2 $ & $ 0.079 $ & $  755.36 $ & $  0.579x10^{15} $ & $ 1.454 $ & $ 1.869 $ \\
$ 14\,\, 14\,\, 47.4  $ &$ - 0\,\, 23\,\, 56.8 $ & $ 0.140 $ & $  612.45 $ & $  0.448x10^{15} $ & $ 1.712 $ & $ 1.515 $ \\
$ 14\,\, 24\,\, 40.4  $ &$ + 2\,\, 44\,\, 46.7 $ & $ 0.055 $ & $  558.59 $ & $  0.208x10^{15} $ & $ 0.954 $ & $ 1.382 $ \\
$ 14\,\, 25\,\, 22.9  $ &$ +63\,\, 11\,\, 22.6 $ & $ 0.139 $ & $ 1052.31 $ & $  0.198x10^{16} $ & $ 2.557 $ & $ 2.604 $ \\
$ 15\,\, 12\,\, 51.1  $ &$ - 1\,\, 28\,\, 47.3 $ & $ 0.122 $ & $  812.48 $ & $  0.895x10^{15} $ & $ 1.943 $ & $ 2.010 $ \\
$ 15\,\, 16\,\, 34.0  $ &$ - 0\,\, 56\,\, 55.7 $ & $ 0.118 $ & $  539.58 $ & $  0.385x10^{15} $ & $ 1.897 $ & $ 1.335 $ \\
$ 15\,\, 29\,\, 12.1  $ &$ +52\,\, 50\,\, 39.8 $ & $ 0.074 $ & $  694.40 $ & $  0.522x10^{15} $ & $ 1.553 $ & $ 1.718 $ \\
$ 16\,\,  1\,\, 22.1  $ &$ +53\,\, 54\,\, 19.1 $ & $ 0.106 $ & $  491.24 $ & $  0.284x10^{15} $ & $ 1.686 $ & $ 1.216 $ \\
$ 16\,\, 11\,\, 17.7  $ &$ +36\,\, 57\,\, 38.2 $ & $ 0.067 $ & $  529.83 $ & $  0.275x10^{15} $ & $ 1.406 $ & $ 1.311 $ \\
$ 16\,\, 56\,\, 20.3  $ &$ +39\,\, 16\,\, 59.9 $ & $ 0.062 $ & $  440.23 $ & $  0.140x10^{15} $ & $ 1.034 $ & $ 1.089 $ \\
$ 16\,\, 59\,\, 45.4  $ &$ +32\,\, 36\,\, 58.0 $ & $ 0.101 $ & $  512.65 $ & $  0.257x10^{15} $ & $ 1.403 $ & $ 1.268 $ \\
$ 17\,\,  2\,\, 42.6  $ &$ +34\,\,  3\,\, 40.7 $ & $ 0.099 $ & $ 1060.16 $ & $  0.161x10^{16} $ & $ 2.059 $ & $ 2.623 $ \\
$ 17\,\, 12\,\, 47.6  $ &$ +64\,\,  3\,\, 47.5 $ & $ 0.080 $ & $ 1037.59 $ & $  0.167x10^{16} $ & $ 2.223 $ & $ 2.567 $ \\
$ 17\,\, 18\,\,  9.9  $ &$ +56\,\, 39\,\, 59.0 $ & $ 0.113 $ & $  676.01 $ & $  0.565x10^{15} $ & $ 1.772 $ & $ 1.673 $ \\
$ 21\,\, 24\,\, 56.3  $ &$ - 6\,\, 56\,\, 47.4 $ & $ 0.118 $ & $  809.90 $ & $  0.102x10^{16} $ & $ 2.224 $ & $ 2.004 $ \\
$ 21\,\, 57\,\, 25.8  $ &$ - 7\,\, 47\,\, 40.6 $ & $ 0.058 $ & $  649.63 $ & $  0.498x10^{15} $ & $ 1.692 $ & $ 1.607 $ \\
$ 22\,\, 16\,\, 15.5  $ &$ - 9\,\, 20\,\, 23.6 $ & $ 0.084 $ & $  488.35 $ & $  0.218x10^{15} $ & $ 1.311 $ & $ 1.208 $ \\
\hline
\end{longtable}
}

\longtab{2}{
\begin{longtable}{llccccc}
\caption{\label{tb:koe} Analogous to table \ref{tb:pop}, for C-K07-I cluster sample.}\\
\hline\hline
$\alpha$ (J$2000$.$00$)& $\delta$ (J$2000$.$00$) & $z$ & $\sigma$ &$M_{vir}$  & $R_{vir}$ & $R_{200}$\\
 $[h\; m \; s]$ & [$\degr\; \arcmin\;\arcsec$]  &   & $[\kms$] & [$h^{-1}\Msol$]  & [$\mpc$] & [$\mpc$] \\
\hline
\endfirsthead
\caption{continued.}\\
\hline\hline
$\alpha$ (J$2000$.$00$)& $\delta$ (J$2000$.$00$) & $z$ & $\sigma$ &$M_{vir}$  & $R_{vir}$ & $R_{200}$ \\
 $[h\; m \; s]$ & [$\degr\; \arcmin\;\arcsec$]  &   & $[\kms$] & [$h^{-1}\Msol$]  & [$\mpc$] & [$\mpc$] \\
\hline
\endhead
\hline
\endfoot
$ 15\,\, 58\,\, 18.4  $ &$ +27\,\, 16\,\, 12.0 $ & $ 0.090 $ & $ 1070.73 $ & $  0.221x10^{16} $ & $ 2.766 $ & $ 2.649 $ \\
$ 15\,\, 10\,\,  1.7  $ &$ +33\,\, 28\,\, 24.2 $ & $ 0.110 $ & $  844.46 $ & $  0.841x10^{15} $ & $ 1.691 $ & $ 2.089 $ \\
$ 13\,\, 25\,\, 16.7  $ &$ +57\,\, 37\,\,  1.2 $ & $ 0.116 $ & $  747.53 $ & $  0.772x10^{15} $ & $ 1.980 $ & $ 1.850 $ \\
$ 10\,\, 19\,\, 55.6  $ &$ +40\,\, 59\,\, 23.3 $ & $ 0.092 $ & $  899.02 $ & $  0.747x10^{15} $ & $ 1.325 $ & $ 2.224 $ \\
$ 16\,\, 42\,\, 38.6  $ &$ +27\,\, 26\,\, 24.0 $ & $ 0.104 $ & $  983.57 $ & $  0.145x10^{16} $ & $ 2.151 $ & $ 2.434 $ \\
$ 14\,\, 28\,\, 30.8  $ &$ +56\,\, 48\,\, 18.7 $ & $ 0.106 $ & $  776.22 $ & $  0.875x10^{15} $ & $ 2.081 $ & $ 1.921 $ \\
$ 10\,\, 27\,\, 46.8  $ &$ +10\,\, 32\,\, 26.9 $ & $ 0.109 $ & $  673.64 $ & $  0.438x10^{15} $ & $ 1.383 $ & $ 1.667 $ \\
$ 15\,\, 39\,\, 50.6  $ &$ +30\,\, 42\,\, 36.0 $ & $ 0.097 $ & $  751.73 $ & $  0.647x10^{15} $ & $ 1.642 $ & $ 1.860 $ \\
$ 14\,\, 35\,\, 20.0  $ &$ +55\,\, 10\,\, 58.1 $ & $ 0.140 $ & $  910.40 $ & $  0.121x10^{16} $ & $ 2.095 $ & $ 2.253 $ \\
$  8\,\, 54\,\, 15.1  $ &$ +29\,\,  3\,\, 30.6 $ & $ 0.085 $ & $  670.67 $ & $  0.409x10^{15} $ & $ 1.305 $ & $ 1.659 $ \\
$ 14\,\, 58\,\, 44.4  $ &$ +47\,\, 32\,\, 24.0 $ & $ 0.085 $ & $  603.92 $ & $  0.450x10^{15} $ & $ 1.767 $ & $ 1.494 $ \\
$  8\,\, 18\,\,  9.8  $ &$ +54\,\, 35\,\, 39.1 $ & $ 0.118 $ & $  438.81 $ & $  0.216x10^{15} $ & $ 1.607 $ & $ 1.086 $ \\
$ 12\,\, 33\,\, 13.9  $ &$ +67\,\,  7\,\, 12.0 $ & $ 0.105 $ & $ 1000.66 $ & $  0.202x10^{16} $ & $ 2.888 $ & $ 2.476 $ \\
$  9\,\,  1\,\, 30.1  $ &$ +55\,\, 39\,\, 16.6 $ & $ 0.116 $ & $  979.65 $ & $  0.547x10^{15} $ & $ 0.817 $ & $ 2.424 $ \\
$  8\,\, 54\,\, 51.5  $ &$ + 0\,\, 50\,\, 50.3 $ & $ 0.108 $ & $  685.18 $ & $  0.525x10^{15} $ & $ 1.604 $ & $ 1.695 $ \\
$ 15\,\, 24\,\, 11.8  $ &$ +29\,\, 51\,\, 22.0 $ & $ 0.114 $ & $ 1219.20 $ & $  0.215x10^{16} $ & $ 2.072 $ & $ 3.017 $ \\
$ 15\,\, 24\,\, 32.0  $ &$ +29\,\, 43\,\, 38.6 $ & $ 0.112 $ & $ 1100.42 $ & $  0.203x10^{16} $ & $ 2.404 $ & $ 2.723 $ \\
$ 15\,\, 47\,\, 41.5  $ &$ +33\,\, 19\,\, 53.0 $ & $ 0.114 $ & $  518.23 $ & $  0.285x10^{15} $ & $ 1.519 $ & $ 1.282 $ \\
$  9\,\, 23\,\, 26.9  $ &$ + 8\,\, 39\,\,  2.5 $ & $ 0.129 $ & $  844.89 $ & $  0.996x10^{15} $ & $ 2.001 $ & $ 2.091 $ \\
$ 11\,\, 53\,\, 51.2  $ &$ +15\,\, 26\,\, 35.9 $ & $ 0.113 $ & $  748.88 $ & $  0.551x10^{15} $ & $ 1.409 $ & $ 1.853 $ \\
$ 16\,\, 20\,\, 31.1  $ &$ +29\,\, 53\,\, 27.6 $ & $ 0.097 $ & $  853.98 $ & $  0.103x10^{16} $ & $ 2.030 $ & $ 2.113 $ \\
$ 16\,\,  3\,\, 19.8  $ &$ +25\,\, 27\,\, 13.3 $ & $ 0.088 $ & $  582.84 $ & $  0.271x10^{15} $ & $ 1.143 $ & $ 1.442 $ \\
$ 14\,\, 54\,\, 37.1  $ &$ +54\,\, 25\,\, 23.2 $ & $ 0.100 $ & $  630.37 $ & $  0.423x10^{15} $ & $ 1.525 $ & $ 1.560 $ \\
$ 21\,\, 30\,\, 27.0  $ &$ - 0\,\,  0\,\, 24.5 $ & $ 0.135 $ & $  555.00 $ & $  0.317x10^{15} $ & $ 1.478 $ & $ 1.373 $ \\
$ 15\,\, 19\,\, 33.7  $ &$ + 4\,\, 20\,\, 16.8 $ & $ 0.103 $ & $  664.20 $ & $  0.361x10^{15} $ & $ 1.173 $ & $ 1.643 $ \\
$ 21\,\, 49\,\,  4.6  $ &$ - 8\,\, 10\,\, 49.4 $ & $ 0.133 $ & $  880.13 $ & $  0.857x10^{15} $ & $ 1.586 $ & $ 2.178 $ \\
$  8\,\, 50\,\,  7.1  $ &$ +29\,\, 32\,\, 52.1 $ & $ 0.104 $ & $  730.39 $ & $  0.707x10^{15} $ & $ 1.899 $ & $ 1.807 $ \\
$  8\,\, 54\,\, 57.5  $ &$ +35\,\, 24\,\, 51.8 $ & $ 0.146 $ & $ 1271.75 $ & $  0.245x10^{16} $ & $ 2.170 $ & $ 3.147 $ \\
$ 14\,\, 29\,\, 21.1  $ &$ +23\,\,  6\,\, 29.9 $ & $ 0.138 $ & $  716.15 $ & $  0.493x10^{15} $ & $ 1.379 $ & $ 1.772 $ \\
$ 11\,\, 40\,\, 33.9  $ &$ +10\,\, 21\,\, 41.0 $ & $ 0.105 $ & $  412.45 $ & $  0.143x10^{15} $ & $ 1.201 $ & $ 1.021 $ \\
$ 11\,\, 13\,\, 48.5  $ &$ - 0\,\, 24\,\, 30.6 $ & $ 0.100 $ & $  561.71 $ & $  0.369x10^{15} $ & $ 1.677 $ & $ 1.390 $ \\
$ 16\,\, 47\,\, 40.6  $ &$ +29\,\, 55\,\, 18.1 $ & $ 0.099 $ & $ 1004.14 $ & $  0.103x10^{16} $ & $ 1.470 $ & $ 2.485 $ \\
$  9\,\,  7\,\, 56.8  $ &$ +52\,\, 48\,\,  2.9 $ & $ 0.099 $ & $  732.91 $ & $  0.462x10^{15} $ & $ 1.232 $ & $ 1.813 $ \\
$ 10\,\,  9\,\, 34.3  $ &$ +44\,\, 42\,\, 38.5 $ & $ 0.146 $ & $ 1080.56 $ & $  0.161x10^{16} $ & $ 1.974 $ & $ 2.674 $ \\
$ 10\,\,  8\,\,  0.4  $ &$ +38\,\,  1\,\,  5.9 $ & $ 0.112 $ & $  574.80 $ & $  0.397x10^{15} $ & $ 1.725 $ & $ 1.422 $ \\
$ 10\,\, 35\,\, 48.9  $ &$ +36\,\,  5\,\, 58.6 $ & $ 0.123 $ & $  462.89 $ & $  0.169x10^{15} $ & $ 1.134 $ & $ 1.145 $ \\
$ 14\,\,  7\,\, 46.5  $ &$ +14\,\,  0\,\, 11.9 $ & $ 0.135 $ & $  658.33 $ & $  0.458x10^{15} $ & $ 1.515 $ & $ 1.629 $ \\
$ 11\,\, 20\,\, 24.7  $ &$ +47\,\,  9\,\, 26.3 $ & $ 0.112 $ & $  605.14 $ & $  0.425x10^{15} $ & $ 1.664 $ & $ 1.497 $ \\
$ 10\,\, 41\,\, 34.6  $ &$ - 0\,\, 36\,\, 53.6 $ & $ 0.135 $ & $  793.10 $ & $  0.753x10^{15} $ & $ 1.717 $ & $ 1.962 $ \\
$  2\,\, 25\,\, 19.2  $ &$ - 8\,\, 44\,\,  5.3 $ & $ 0.054 $ & $  479.32 $ & $  0.131x10^{15} $ & $ 0.817 $ & $ 1.186 $ \\
$  0\,\, 28\,\, 22.5  $ &$ +13\,\, 52\,\, 23.5 $ & $ 0.141 $ & $  578.78 $ & $  0.496x10^{15} $ & $ 2.125 $ & $ 1.432 $ \\
$ 12\,\, 12\,\, 52.5  $ &$ + 6\,\,  3\,\, 48.2 $ & $ 0.137 $ & $  580.40 $ & $  0.353x10^{15} $ & $ 1.502 $ & $ 1.436 $ \\
$ 12\,\, 47\,\, 20.0  $ &$ + 0\,\,  8\,\, 39.1 $ & $ 0.089 $ & $  779.99 $ & $  0.680x10^{15} $ & $ 1.604 $ & $ 1.930 $ \\
$ 14\,\, 47\,\, 33.3  $ &$ +33\,\,  2\,\, 38.4 $ & $ 0.087 $ & $  314.85 $ & $  0.901x10^{14} $ & $ 1.303 $ & $ 0.779 $ \\
$ 15\,\, 29\,\,  1.3  $ &$ +30\,\, 46\,\, 20.6 $ & $ 0.105 $ & $  647.03 $ & $  0.413x10^{15} $ & $ 1.414 $ & $ 1.601 $ \\
$  8\,\, 29\,\, 29.4  $ &$ +39\,\,  9\,\, 27.7 $ & $ 0.092 $ & $  659.36 $ & $  0.417x10^{15} $ & $ 1.376 $ & $ 1.632 $ \\
$  1\,\,  1\,\,  3.3  $ &$ - 9\,\, 57\,\, 34.6 $ & $ 0.150 $ & $  776.20 $ & $  0.765x10^{15} $ & $ 1.820 $ & $ 1.921 $ \\
$ 13\,\, 21\,\, 54.0  $ &$ +57\,\, 32\,\, 21.5 $ & $ 0.118 $ & $  379.97 $ & $  0.121x10^{15} $ & $ 1.201 $ & $ 0.940 $ \\
$ 20\,\, 51\,\, 42.2  $ &$ - 0\,\,  4\,\,  8.0 $ & $ 0.108 $ & $  468.59 $ & $  0.180x10^{15} $ & $ 1.177 $ & $ 1.159 $ \\
$ 21\,\, 26\,\, 36.9  $ &$ - 6\,\, 39\,\, 13.0 $ & $ 0.124 $ & $  513.26 $ & $  0.291x10^{15} $ & $ 1.584 $ & $ 1.270 $ \\
$ 12\,\,  4\,\,  8.3  $ &$ + 4\,\, 19\,\, 33.2 $ & $ 0.136 $ & $  737.57 $ & $  0.654x10^{15} $ & $ 1.724 $ & $ 1.825 $ \\
$ 15\,\, 16\,\, 46.1  $ &$ - 0\,\, 54\,\,  2.9 $ & $ 0.118 $ & $  877.76 $ & $  0.118x10^{16} $ & $ 2.204 $ & $ 2.172 $ \\
$ 10\,\, 58\,\,  1.5  $ &$ +41\,\,  2\,\, 47.8 $ & $ 0.129 $ & $  919.45 $ & $  0.867x10^{15} $ & $ 1.471 $ & $ 2.275 $ \\
$ 11\,\, 21\,\, 26.4  $ &$ +53\,\, 44\,\, 56.8 $ & $ 0.104 $ & $  694.49 $ & $  0.536x10^{15} $ & $ 1.592 $ & $ 1.718 $ \\
$  8\,\, 14\,\, 52.7  $ &$ + 6\,\, 57\,\, 42.8 $ & $ 0.110 $ & $  575.06 $ & $  0.283x10^{15} $ & $ 1.225 $ & $ 1.423 $ \\
$ 12\,\, 48\,\, 40.0  $ &$ +62\,\, 37\,\,  3.0 $ & $ 0.104 $ & $  640.81 $ & $  0.294x10^{15} $ & $ 1.025 $ & $ 1.586 $ \\
$ 16\,\, 45\,\, 37.9  $ &$ +27\,\, 18\,\, 45.0 $ & $ 0.101 $ & $  511.66 $ & $  0.188x10^{15} $ & $ 1.029 $ & $ 1.266 $ \\
$  9\,\, 15\,\, 57.8  $ &$ + 5\,\, 29\,\,  0.6 $ & $ 0.142 $ & $  624.91 $ & $  0.445x10^{15} $ & $ 1.633 $ & $ 1.546 $ \\
$ 10\,\, 20\,\, 15.3  $ &$ +39\,\, 18\,\, 13.3 $ & $ 0.147 $ & $  445.84 $ & $  0.169x10^{15} $ & $ 1.218 $ & $ 1.103 $ \\
$ 21\,\, 47\,\,  8.5  $ &$ - 7\,\, 53\,\, 30.8 $ & $ 0.130 $ & $  761.77 $ & $  0.447x10^{15} $ & $ 1.104 $ & $ 1.885 $ \\
$ 21\,\, 41\,\, 16.2  $ &$ - 8\,\, 23\,\,  9.2 $ & $ 0.129 $ & $  826.61 $ & $  0.702x10^{15} $ & $ 1.474 $ & $ 2.045 $ \\
$ 15\,\, 10\,\,  8.9  $ &$ + 2\,\, 59\,\, 48.8 $ & $ 0.093 $ & $  571.04 $ & $  0.320x10^{15} $ & $ 1.408 $ & $ 1.413 $ \\
$ 11\,\, 46\,\, 16.8  $ &$ +11\,\, 11\,\,  1.7 $ & $ 0.112 $ & $  397.07 $ & $  0.147x10^{15} $ & $ 1.337 $ & $ 0.983 $ \\
$ 22\,\, 58\,\,  2.6  $ &$ +14\,\,  9\,\, 48.6 $ & $ 0.094 $ & $  433.07 $ & $  0.103x10^{15} $ & $ 0.785 $ & $ 1.072 $ \\
$ 15\,\, 43\,\, 12.4  $ &$ + 5\,\, 52\,\,  6.2 $ & $ 0.106 $ & $  835.13 $ & $  0.932x10^{15} $ & $ 1.916 $ & $ 2.066 $ \\
$ 11\,\, 42\,\, 17.5  $ &$ +10\,\, 17\,\, 30.8 $ & $ 0.117 $ & $  435.71 $ & $  0.134x10^{15} $ & $ 1.013 $ & $ 1.078 $ \\
$  9\,\,  6\,\, 37.6  $ &$ +10\,\, 19\,\,  5.2 $ & $ 0.134 $ & $  762.70 $ & $  0.782x10^{15} $ & $ 1.928 $ & $ 1.887 $ \\
$ 11\,\, 44\,\, 53.8  $ &$ +51\,\, 16\,\,  1.2 $ & $ 0.129 $ & $  804.40 $ & $  0.881x10^{15} $ & $ 1.953 $ & $ 1.990 $ \\
$ 12\,\, 55\,\, 58.8  $ &$ +62\,\,  8\,\, 48.5 $ & $ 0.105 $ & $  653.23 $ & $  0.303x10^{15} $ & $ 1.018 $ & $ 1.616 $ \\
$  0\,\, 21\,\,  7.6  $ &$ - 0\,\, 55\,\, 31.4 $ & $ 0.111 $ & $  623.97 $ & $  0.375x10^{15} $ & $ 1.382 $ & $ 1.544 $ \\
$ 12\,\, 15\,\,  1.9  $ &$ - 2\,\, 41\,\, 25.8 $ & $ 0.110 $ & $  347.44 $ & $  0.823x10^{14} $ & $ 0.978 $ & $ 0.860 $ \\
$ 10\,\, 29\,\, 25.1  $ &$ +37\,\, 37\,\, 48.0 $ & $ 0.108 $ & $  304.95 $ & $  0.108x10^{15} $ & $ 1.666 $ & $ 0.755 $ \\
$ 16\,\, 47\,\, 44.9  $ &$ +29\,\, 10\,\, 55.2 $ & $ 0.134 $ & $ 1077.76 $ & $  0.179x10^{16} $ & $ 2.205 $ & $ 2.667 $ \\
$ 15\,\, 23\,\, 36.5  $ &$ +31\,\,  1\,\, 17.4 $ & $ 0.074 $ & $  909.39 $ & $  0.120x10^{16} $ & $ 2.082 $ & $ 2.250 $ \\
$ 22\,\,  9\,\, 56.2  $ &$ - 7\,\, 50\,\, 50.6 $ & $ 0.116 $ & $  555.93 $ & $  0.355x10^{15} $ & $ 1.647 $ & $ 1.376 $ \\
$ 11\,\, 53\,\, 58.9  $ &$ + 9\,\, 39\,\, 29.9 $ & $ 0.103 $ & $  766.07 $ & $  0.557x10^{15} $ & $ 1.360 $ & $ 1.896 $ \\
$  7\,\, 59\,\, 39.4  $ &$ +41\,\, 50\,\, 24.0 $ & $ 0.132 $ & $  647.49 $ & $  0.477x10^{15} $ & $ 1.631 $ & $ 1.602 $ \\
$ 15\,\, 39\,\, 56.0  $ &$ - 2\,\, 11\,\, 47.8 $ & $ 0.150 $ & $  941.81 $ & $  0.129x10^{16} $ & $ 2.088 $ & $ 2.330 $ \\
$ 11\,\, 43\,\, 45.5  $ &$ +46\,\, 20\,\, 56.4 $ & $ 0.114 $ & $  496.34 $ & $  0.290x10^{15} $ & $ 1.689 $ & $ 1.228 $ \\
$ 14\,\, 29\,\, 14.4  $ &$ +33\,\, 59\,\, 23.6 $ & $ 0.130 $ & $  919.26 $ & $  0.147x10^{16} $ & $ 2.500 $ & $ 2.275 $ \\
$ 11\,\, 48\,\, 59.9  $ &$ +51\,\, 36\,\, 54.4 $ & $ 0.131 $ & $  883.70 $ & $  0.114x10^{16} $ & $ 2.095 $ & $ 2.187 $ \\
$ 15\,\, 16\,\, 39.5  $ &$ + 2\,\, 47\,\, 12.8 $ & $ 0.111 $ & $  466.11 $ & $  0.173x10^{15} $ & $ 1.140 $ & $ 1.153 $ \\
$ 14\,\, 14\,\, 12.3  $ &$ + 6\,\, 52\,\,  3.0 $ & $ 0.109 $ & $  639.74 $ & $  0.469x10^{15} $ & $ 1.642 $ & $ 1.583 $ \\
$ 21\,\, 40\,\,  1.0  $ &$ - 8\,\,  5\,\, 53.5 $ & $ 0.132 $ & $  709.70 $ & $  0.653x10^{15} $ & $ 1.860 $ & $ 1.756 $ \\
$ 10\,\, 49\,\, 37.8  $ &$ + 3\,\, 38\,\, 51.4 $ & $ 0.151 $ & $  759.96 $ & $  0.816x10^{15} $ & $ 2.025 $ & $ 1.880 $ \\
$ 23\,\, 49\,\, 28.7  $ &$ +15\,\, 14\,\,  6.7 $ & $ 0.114 $ & $  645.35 $ & $  0.361x10^{15} $ & $ 1.243 $ & $ 1.597 $ \\
$ 11\,\, 44\,\,  1.4  $ &$ - 1\,\, 45\,\, 27.7 $ & $ 0.106 $ & $  647.10 $ & $  0.400x10^{15} $ & $ 1.370 $ & $ 1.601 $ \\
$ 13\,\, 40\,\, 57.8  $ &$ + 3\,\,  9\,\, 53.3 $ & $ 0.115 $ & $  483.52 $ & $  0.270x10^{15} $ & $ 1.653 $ & $ 1.196 $ \\
$ 16\,\,  7\,\, 33.4  $ &$ +23\,\, 13\,\, 18.1 $ & $ 0.088 $ & $  420.68 $ & $  0.132x10^{15} $ & $ 1.069 $ & $ 1.041 $ \\
$ 15\,\, 42\,\, 34.4  $ &$ +41\,\, 49\,\, 13.1 $ & $ 0.141 $ & $  701.91 $ & $  0.616x10^{15} $ & $ 1.793 $ & $ 1.737 $ \\
$ 14\,\, 30\,\, 33.6  $ &$ +24\,\, 40\,\,  3.4 $ & $ 0.134 $ & $  873.70 $ & $  0.123x10^{16} $ & $ 2.306 $ & $ 2.162 $ \\
$ 10\,\, 30\,\, 17.4  $ &$ +41\,\,  8\,\, 28.7 $ & $ 0.091 $ & $  483.52 $ & $  0.180x10^{15} $ & $ 1.105 $ & $ 1.196 $ \\
$  8\,\, 54\,\, 10.1  $ &$ +23\,\, 34\,\, 33.2 $ & $ 0.112 $ & $  481.69 $ & $  0.222x10^{15} $ & $ 1.372 $ & $ 1.192 $ \\
$  7\,\, 53\,\, 25.2  $ &$ +34\,\, 16\,\, 32.5 $ & $ 0.139 $ & $  410.75 $ & $  0.151x10^{15} $ & $ 1.286 $ & $ 1.016 $ \\
$  8\,\, 39\,\, 16.7  $ &$ + 8\,\, 22\,\, 50.9 $ & $ 0.132 $ & $  815.42 $ & $  0.996x10^{15} $ & $ 2.148 $ & $ 2.018 $ \\
$ 16\,\, 34\,\,  4.8  $ &$ +40\,\, 55\,\, 59.9 $ & $ 0.136 $ & $  591.23 $ & $  0.323x10^{15} $ & $ 1.326 $ & $ 1.463 $ \\
$ 10\,\, 31\,\, 35.7  $ &$ +35\,\,  3\,\, 15.8 $ & $ 0.122 $ & $  928.83 $ & $  0.130x10^{16} $ & $ 2.161 $ & $ 2.298 $ \\
$  9\,\, 51\,\, 14.2  $ &$ + 8\,\, 16\,\, 51.6 $ & $ 0.142 $ & $  737.42 $ & $  0.729x10^{15} $ & $ 1.921 $ & $ 1.825 $ \\
$ 10\,\, 24\,\, 37.1  $ &$ +50\,\, 13\,\, 28.9 $ & $ 0.156 $ & $  487.80 $ & $  0.378x10^{15} $ & $ 2.275 $ & $ 1.207 $ \\
$ 10\,\, 36\,\, 52.7  $ &$ +44\,\, 52\,\, 35.0 $ & $ 0.124 $ & $  763.93 $ & $  0.593x10^{15} $ & $ 1.457 $ & $ 1.890 $ \\
$ 13\,\, 56\,\, 44.8  $ &$ +44\,\, 54\,\, 11.9 $ & $ 0.126 $ & $  717.58 $ & $  0.711x10^{15} $ & $ 1.981 $ & $ 1.776 $ \\
$ 10\,\, 16\,\, 22.8  $ &$ +33\,\, 38\,\, 17.5 $ & $ 0.129 $ & $  847.43 $ & $  0.110x10^{16} $ & $ 2.203 $ & $ 2.097 $ \\
$ 14\,\,  8\,\,  6.7  $ &$ + 6\,\, 33\,\, 34.9 $ & $ 0.112 $ & $  601.25 $ & $  0.385x10^{15} $ & $ 1.528 $ & $ 1.488 $ \\
$ 11\,\, 27\,\, 30.3  $ &$ + 0\,\,  9\,\, 19.4 $ & $ 0.131 $ & $  517.61 $ & $  0.194x10^{15} $ & $ 1.039 $ & $ 1.281 $ \\
$  8\,\, 44\,\, 47.3  $ &$ +27\,\, 41\,\, 20.8 $ & $ 0.085 $ & $  483.87 $ & $  0.160x10^{15} $ & $ 0.983 $ & $ 1.197 $ \\
$ 11\,\, 12\,\,  6.7  $ &$ +30\,\, 40\,\, 41.9 $ & $ 0.074 $ & $  623.97 $ & $  0.326x10^{15} $ & $ 1.202 $ & $ 1.544 $ \\
$ 21\,\, 50\,\, 36.1  $ &$ - 7\,\, 53\,\, 28.0 $ & $ 0.122 $ & $  639.67 $ & $  0.351x10^{15} $ & $ 1.229 $ & $ 1.583 $ \\
$ 13\,\, 14\,\, 24.7  $ &$ +62\,\, 19\,\, 45.8 $ & $ 0.135 $ & $  580.22 $ & $  0.341x10^{15} $ & $ 1.452 $ & $ 1.436 $ \\
$ 11\,\, 43\,\,  4.0  $ &$ +11\,\,  1\,\, 36.5 $ & $ 0.153 $ & $  679.73 $ & $  0.363x10^{15} $ & $ 1.128 $ & $ 1.682 $ \\
$ 12\,\, 27\,\, 32.3  $ &$ +49\,\, 28\,\, 44.0 $ & $ 0.119 $ & $  560.16 $ & $  0.261x10^{15} $ & $ 1.194 $ & $ 1.386 $ \\
$ 10\,\, 32\,\,  9.4  $ &$ +53\,\, 19\,\, 12.7 $ & $ 0.135 $ & $  557.49 $ & $  0.330x10^{15} $ & $ 1.520 $ & $ 1.379 $ \\
$  9\,\, 24\,\, 30.8  $ &$ + 7\,\, 56\,\, 56.0 $ & $ 0.105 $ & $  423.20 $ & $  0.137x10^{15} $ & $ 1.096 $ & $ 1.047 $ \\
$ 11\,\,  4\,\, 24.8  $ &$ +48\,\, 36\,\, 42.8 $ & $ 0.111 $ & $  581.89 $ & $  0.337x10^{15} $ & $ 1.425 $ & $ 1.440 $ \\
$ 14\,\, 37\,\, 50.0  $ &$ +48\,\, 36\,\, 30.6 $ & $ 0.122 $ & $  782.47 $ & $  0.769x10^{15} $ & $ 1.801 $ & $ 1.936 $ \\
$  8\,\, 43\,\, 33.9  $ &$ +38\,\, 55\,\, 56.3 $ & $ 0.121 $ & $  484.42 $ & $  0.218x10^{15} $ & $ 1.330 $ & $ 1.199 $ \\
$ 13\,\, 21\,\, 18.9  $ &$ - 0\,\, 43\,\, 43.7 $ & $ 0.108 $ & $  627.17 $ & $  0.271x10^{15} $ & $ 0.987 $ & $ 1.552 $ \\
$ 12\,\,  1\,\, 39.0  $ &$ +58\,\,  1\,\, 38.6 $ & $ 0.104 $ & $  806.22 $ & $  0.897x10^{15} $ & $ 1.980 $ & $ 1.995 $ \\
$ 10\,\, 45\,\, 58.5  $ &$ + 1\,\, 26\,\, 56.0 $ & $ 0.105 $ & $  375.80 $ & $  0.143x10^{15} $ & $ 1.451 $ & $ 0.930 $ \\
$ 10\,\, 45\,\,  1.5  $ &$ +58\,\,  5\,\,  6.0 $ & $ 0.116 $ & $  512.04 $ & $  0.226x10^{15} $ & $ 1.239 $ & $ 1.267 $ \\
$ 16\,\, 24\,\, 56.8  $ &$ +28\,\, 33\,\, 41.4 $ & $ 0.145 $ & $  887.74 $ & $  0.724x10^{15} $ & $ 1.317 $ & $ 2.197 $ \\
$  9\,\, 46\,\,  8.6  $ &$ + 3\,\, 46\,\, 40.8 $ & $ 0.119 $ & $  545.97 $ & $  0.317x10^{15} $ & $ 1.523 $ & $ 1.351 $ \\
$ 16\,\, 47\,\, 46.2  $ &$ +29\,\, 10\,\,  9.5 $ & $ 0.134 $ & $  655.71 $ & $  0.600x10^{15} $ & $ 2.001 $ & $ 1.622 $ \\
$ 11\,\, 37\,\, 18.2  $ &$ +57\,\,  8\,\,  2.8 $ & $ 0.117 $ & $  560.11 $ & $  0.201x10^{15} $ & $ 0.9a commentary 20 $ & $ 1.386 $ \\
$ 13\,\, 48\,\, 13.1  $ &$ +57\,\, 41\,\, 38.0 $ & $ 0.127 $ & $ 1090.14 $ & $  0.193x10^{16} $ & $ 2.323 $ & $ 2.697 $ \\
$  0\,\, 45\,\, 49.8  $ &$ - 0\,\, 51\,\,  1.4 $ & $ 0.105 $ & $  567.44 $ & $  0.311x10^{15} $ & $ 1.386 $ & $ 1.404 $ \\
$ 12\,\,  1\,\,  4.5  $ &$ +15\,\, 12\,\, 35.6 $ & $ 0.109 $ & $  529.12 $ & $  0.179x10^{15} $ & $ 0.915 $ & $ 1.309 $ \\
$ 14\,\, 27\,\, 24.5  $ &$ +55\,\, 45\,\,  1.1 $ & $ 0.131 $ & $  713.57 $ & $  0.624x10^{15} $ & $ 1.757 $ & $ 1.766 $ \\
$ 10\,\,  7\,\, 51.3  $ &$ +62\,\, 30\,\,  2.2 $ & $ 0.137 $ & $  560.11 $ & $  0.187x10^{15} $ & $ 0.853 $ & $ 1.386 $ \\
$ 16\,\, 41\,\, 40.6  $ &$ +22\,\,  0\,\, 54.4 $ & $ 0.151 $ & $  850.87 $ & $  0.913x10^{15} $ & $ 1.807 $ & $ 2.105 $ \\
$ 13\,\, 23\,\, 36.1  $ &$ + 4\,\, 42\,\, 40.3 $ & $ 0.134 $ & $  794.57 $ & $  0.724x10^{15} $ & $ 1.643 $ & $ 1.966 $ \\
$  8\,\, 56\,\,  0.8  $ &$ +48\,\, 29\,\, 10.3 $ & $ 0.124 $ & $  668.77 $ & $  0.421x10^{15} $ & $ 1.351 $ & $ 1.655 $ \\
$ 15\,\, 21\,\, 15.7  $ &$ +32\,\,  5\,\, 12.1 $ & $ 0.111 $ & $  827.47 $ & $  0.826x10^{15} $ & $ 1.729 $ & $ 2.047 $ \\
$ 13\,\, 59\,\, 20.8  $ &$ +49\,\, 26\,\, 51.4 $ & $ 0.106 $ & $  553.87 $ & $  0.289x10^{15} $ & $ 1.353 $ & $ 1.370 $ \\
$  9\,\, 26\,\, 59.7  $ &$ +54\,\, 22\,\, 36.1 $ & $ 0.125 $ & $  673.39 $ & $  0.276x10^{15} $ & $ 0.872 $ & $ 1.666 $ \\
$ 15\,\, 36\,\, 29.8  $ &$ - 1\,\, 57\,\, 48.2 $ & $ 0.145 $ & $  716.22 $ & $  0.619x10^{15} $ & $ 1.730 $ & $ 1.772 $ \\
$ 11\,\, 21\,\, 35.4  $ &$ +35\,\, 23\,\, 24.7 $ & $ 0.103 $ & $  591.23 $ & $  0.395x10^{15} $ & $ 1.619 $ & $ 1.463 $ \\
$ 11\,\, 48\,\, 20.4  $ &$ +10\,\, 21\,\,  8.3 $ & $ 0.113 $ & $  999.27 $ & $  0.158x10^{16} $ & $ 2.269 $ & $ 2.473 $ \\
$ 10\,\, 48\,\,  8.8  $ &$ +31\,\, 28\,\, 31.1 $ & $ 0.115 $ & $  623.88 $ & $  0.302x10^{15} $ & $ 1.112 $ & $ 1.544 $ \\
$ 14\,\, 23\,\, 49.9  $ &$ + 6\,\, 14\,\, 32.3 $ & $ 0.113 $ & $  463.77 $ & $  0.224x10^{15} $ & $ 1.495 $ & $ 1.148 $ \\
$ 14\,\, 48\,\, 18.4  $ &$ + 3\,\, 31\,\, 44.4 $ & $ 0.124 $ & $  741.07 $ & $  0.522x10^{15} $ & $ 1.362 $ & $ 1.834 $ \\
$  8\,\,  1\,\,  7.0  $ &$ +17\,\, 58\,\, 45.1 $ & $ 0.144 $ & $  454.09 $ & $  0.109x10^{15} $ & $ 0.756 $ & $ 1.124 $ \\
$ 11\,\, 13\,\, 33.7  $ &$ +37\,\, 22\,\, 27.8 $ & $ 0.102 $ & $  556.72 $ & $  0.300x10^{15} $ & $ 1.386 $ & $ 1.378 $ \\
$  9\,\, 40\,\, 30.1  $ &$ + 2\,\, 28\,\, 35.4 $ & $ 0.151 $ & $  865.54 $ & $  0.813x10^{15} $ & $ 1.556 $ & $ 2.142 $ \\
$  9\,\,  1\,\, 18.4  $ &$ +58\,\, 16\,\,  8.0 $ & $ 0.097 $ & $  839.35 $ & $  0.547x10^{15} $ & $ 1.113 $ & $ 2.077 $ \\
$ 14\,\, 42\,\, 58.5  $ &$ +55\,\, 10\,\, 55.2 $ & $ 0.105 $ & $  435.71 $ & $  0.111x10^{15} $ & $ 0.840 $ & $ 1.078 $ \\
$  8\,\, 44\,\, 36.5  $ &$ +29\,\, 21\,\, 12.6 $ & $ 0.099 $ & $  505.37 $ & $  0.171x10^{15} $ & $ 0.960 $ & $ 1.250 $ \\
$ 10\,\, 38\,\,  1.8  $ &$ +41\,\, 46\,\, 25.7 $ & $ 0.124 $ & $  567.73 $ & $  0.289x10^{15} $ & $ 1.285 $ & $ 1.405 $ \\
$ 14\,\, 52\,\, 33.6  $ &$ +50\,\, 55\,\, 22.1 $ & $ 0.131 $ & $  560.11 $ & $  0.313x10^{15} $ & $ 1.430 $ & $ 1.386 $ \\
$ 14\,\,  6\,\, 25.0  $ &$ + 6\,\, 35\,\,  3.1 $ & $ 0.113 $ & $  477.41 $ & $  0.239x10^{15} $ & $ 1.501 $ & $ 1.181 $ \\
$ 12\,\, 16\,\, 49.1  $ &$ - 3\,\,  7\,\, 55.2 $ & $ 0.111 $ & $  435.64 $ & $  0.134x10^{15} $ & $ 1.012 $ & $ 1.078 $ \\
$ 23\,\,  6\,\, 18.9  $ &$ +14\,\,  9\,\, 40.7 $ & $ 0.112 $ & $  475.07 $ & $  0.236x10^{15} $ & $ 1.500 $ & $ 1.17●5 $ \\
$  1\,\, 42\,\, 25.6  $ &$ -10\,\, 16\,\, 11.6 $ & $ 0.112 $ & $  507.75 $ & $  0.250x10^{15} $ & $ 1.391 $ & $ 1.256 $ \\
$ 14\,\, 17\,\, 54.2  $ &$ +43\,\, 23\,\, 17.2 $ & $ 0.105 $ & $  430.73 $ & $  0.168x10^{15} $ & $ 1.296 $ & $ 1.066 $ \\
$ 13\,\, 23\,\, 48.6  $ &$ + 1\,\,  6\,\, 48.2 $ & $ 0.108 $ & $  668.18 $ & $  0.369x10^{15} $ & $ 1.187 $ & $ 1.653 $ \\
$  8\,\, 44\,\, 21.1  $ &$ +51\,\, 24\,\, 21.2 $ & $ 0.097 $ & $  496.13 $ & $  0.186x10^{15} $ & $ 1.081 $ & $ 1.228 $ \\
$ 11\,\, 51\,\, 10.7  $ &$ - 3\,\,  1\,\, 41.2 $ & $ 0.091 $ & $  212.72 $ & $  0.268x10^{14} $ & $ 0.850 $ & $ 0.52●6 $ \\
$ 10\,\, 19\,\, 35.7  $ &$ +14\,\,  2\,\, 26.5 $ & $ 0.146 $ & $  631.06 $ & $  0.544x10^{15} $ & $ 1.959 $ & $ 1.561 $ \\
$ 12\,\, 13\,\, 58.5  $ &$ +63\,\, 13\,\, 16.0 $ & $ 0.133 $ & $  541.44 $ & $  0.170x10^{15} $ & $ 0.831 $ & $ 1.340 $ \\
$ 16\,\, 20\,\,  8.2  $ &$ +42\,\, 30\,\,  2.9 $ & $ 0.135 $ & $  658.63 $ & $  0.524x10^{15} $ & $ 1.732 $ & $ 1.630 $ \\
$ 10\,\, 17\,\, 17.7  $ &$ + 8\,\, 40\,\, 57.7 $ & $ 0.104 $ & $  522.25 $ & $  0.215x10^{15} $ & $ 1.131 $ & $ 1.292 $ \\
$ 11\,\, 37\,\, 36.7  $ &$ +32\,\, 26\,\, 40.6 $ & $ 0.103 $ & $  510.52 $ & $  0.174x10^{15} $ & $ 0.959 $ & $ 1.263 $ \\
$  8\,\, 28\,\,  0.9  $ &$ +28\,\, 15\,\, 51.8 $ & $ 0.093 $ & $  418.11 $ & $  0.136x10^{15} $ & $ 1.119 $ & $ 1.035 $ \\
$  9\,\, 16\,\, 26.7  $ &$ + 5\,\, 55\,\, 15.2 $ & $ 0.134 $ & $  754.05 $ & $  0.637x10^{15} $ & $ 1.606 $ & $ 1.866 $ \\
$ 10\,\,  9\,\, 35.2  $ &$ + 7\,\,  9\,\, 50.8 $ & $ 0.099 $ & $  425.58 $ & $  0.131x10^{15} $ & $ 1.038 $ & $ 1.053 $ \\
$ 11\,\, 40\,\,  9.3  $ &$ +32\,\, 23\,\, 20.4 $ & $ 0.131 $ & $  818.31 $ & $  0.852x10^{15} $ & $ 1.825 $ & $ 2.025 $ \\
$ 23\,\, 47\,\, 55.1  $ &$ +14\,\, 54\,\, 52.2 $ & $ 0.105 $ & $  424.17 $ & $  0.205x10^{15} $ & $ 1.630 $ & $ 1.050 $ \\
$ 10\,\, 28\,\, 28.7  $ &$ + 9\,\, 38\,\, 58.6 $ & $ 0.106 $ & $  581.90 $ & $  0.382x10^{15} $ & $ 1.619 $ & $ 1.440 $ \\
$  0\,\,  5\,\, 21.1  $ &$ +15\,\, 58\,\,  4.8 $ & $ 0.117 $ & $  541.07 $ & $  0.430x10^{15} $ & $ 2.106 $ & $ 1.339 $ \\
$  9\,\,  6\,\, 49.9  $ &$ + 4\,\, 46\,\, 45.8 $ & $ 0.126 $ & $  727.16 $ & $  0.475x10^{15} $ & $ 1.288 $ & $ 1.799 $ \\
$ 10\,\, 11\,\, 58.3  $ &$ +33\,\, 34\,\, 15.2 $ & $ 0.127 $ & $  631.29 $ & $  0.349x10^{15} $ & $ 1.255 $ & $ 1.562 $ \\
$  9\,\, 39\,\,  9.5  $ &$ +37\,\, 40\,\,  1.6 $ & $ 0.148 $ & $  835.21 $ & $  0.588x10^{15} $ & $ 1.208 $ & $ 2.067 $ \\
$ 10\,\, 43\,\, 52.0  $ &$ + 1\,\,  3\,\, 42.1 $ & $ 0.117 $ & $  419.59 $ & $  0.140x10^{15} $ & $ 1.137 $ & $ 1.038 $ \\
$ 13\,\, 11\,\, 22.6  $ &$ +48\,\, 30\,\, 45.4 $ & $ 0.140 $ & $  807.27 $ & $  0.970x10^{15} $ & $ 2.134 $ & $ 1.997 $ \\
$  3\,\,  5\,\, 59.3  $ &$ - 0\,\,  9\,\, 59.4 $ & $ 0.110 $ & $  623.76 $ & $  0.256x10^{15} $ & $ 0.944 $ & $ 1.543 $ \\
$ 12\,\, 10\,\, 52.2  $ &$ +15\,\, 55\,\, 35.0 $ & $ 0.107 $ & $  553.89 $ & $  0.262x10^{15} $ & $ 1.227 $ & $ 1.371 $ \\
$ 12\,\, 19\,\,  9.3  $ &$ +63\,\, 31\,\, 42.2 $ & $ 0.107 $ & $  611.17 $ & $  0.405x10^{15} $ & $ 1.554 $ & $ 1.512 $ \\
$ 10\,\, 27\,\, 26.5  $ &$ +37\,\, 52\,\, 37.6 $ & $ 0.107 $ & $  403.63 $ & $  0.163x10^{15} $ & $ 1.435 $ & $ 0.999 $ \\
$  9\,\, 50\,\, 53.3  $ &$ +28\,\, 48\,\, 15.5 $ & $ 0.113 $ & $  810.78 $ & $  0.837x10^{15} $ & $ 1.827 $ & $ 2.006 $ \\
$ 12\,\, 11\,\,  9.8  $ &$ + 6\,\, 10\,\, 49.8 $ & $ 0.138 $ & $  559.20 $ & $  0.240x10^{15} $ & $ 1.101 $ & $ 1.384 $ \\
$ 10\,\, 25\,\,  0.7  $ &$ +49\,\, 51\,\, 33.5 $ & $ 0.135 $ & $ 1120.42 $ & $  0.153x10^{16} $ & $ 1.743 $ & $ 2.772 $ \\
$ 14\,\, 33\,\, 39.1  $ &$ +61\,\, 22\,\,  3.4 $ & $ 0.113 $ & $  618.08 $ & $  0.446x10^{15} $ & $ 1.675 $ & $ 1.529 $ \\
$ 11\,\, 41\,\, 15.1  $ &$ +10\,\, 43\,\, 15.6 $ & $ 0.105 $ & $  636.26 $ & $  0.436x10^{15} $ & $ 1.545 $ & $ 1.574 $ \\
$  9\,\, 53\,\, 52.6  $ &$ +12\,\, 15\,\, 55.4 $ & $ 0.129 $ & $  521.13 $ & $  0.286x10^{15} $ & $ 1.508 $ & $ 1.289 $ \\
$ 10\,\, 28\,\, 45.1  $ &$ +37\,\, 52\,\, 38.3 $ & $ 0.107 $ & $  541.02 $ & $  0.421x10^{15} $ & $ 2.062 $ & $ 1.339 $ \\
$ 10\,\, 10\,\, 45.3  $ &$ +33\,\, 43\,\,  3.7 $ & $ 0.135 $ & $  664.31 $ & $  0.685x10^{15} $ & $ 2.224 $ & $ 1.644 $ \\
$ 10\,\, 21\,\, 16.9  $ &$ +12\,\, 13\,\,  0.1 $ & $ 0.129 $ & $  537.06 $ & $  0.341x10^{15} $ & $ 1.695 $ & $ 1.329 $ \\
$  9\,\, 46\,\, 38.1  $ &$ +29\,\, 42\,\,  2.2 $ & $ 0.112 $ & $  253.32 $ & $  0.473x10^{14} $ & $ 1.057 $ & $ 0.627 $ \\
$ 13\,\,  2\,\, 59.1  $ &$ +51\,\, 18\,\, 56.5 $ & $ 0.121 $ & $  554.46 $ & $  0.284x10^{15} $ & $ 1.323 $ & $ 1.372 $ \\
$  1\,\, 19\,\, 16.5  $ &$ +14\,\, 42\,\, 31.0 $ & $ 0.129 $ & $  525.18 $ & $  0.339x10^{15} $ & $ 1.760 $ & $ 1.299 $ \\
$  8\,\, 25\,\, 48.4  $ &$ +56\,\,  0\,\, 48.2 $ & $ 0.138 $ & $  648.50 $ & $  0.350x10^{15} $ & $ 1.195 $ & $ 1.605 $ \\
$ 13\,\, 48\,\, 19.9  $ &$ +57\,\, 45\,\, 34.6 $ & $ 0.127 $ & $  541.44 $ & $  0.323x10^{15} $ & $ 1.582 $ & $ 1.340 $ \\
$ 22\,\, 56\,\, 28.3  $ &$ - 0\,\, 32\,\, 53.9 $ & $ 0.110 $ & $  430.58 $ & $  0.123x10^{15} $ & $ 0.948 $ & $ 1.065 $ \\
$  7\,\, 59\,\, 41.8  $ &$ +28\,\, 46\,\, 55.6 $ & $ 0.139 $ & $  503.43 $ & $  0.177x10^{15} $ & $ 0.999 $ & $ 1.246 $ \\
$ 14\,\, 32\,\, 22.8  $ &$ +47\,\,  6\,\, 38.2 $ & $ 0.109 $ & $  311.13 $ & $  0.698x10^{14} $ & $ 1.034 $ & $ 0.770 $ \\
$ 11\,\, 15\,\, 16.5  $ &$ +53\,\, 42\,\, 42.5 $ & $ 0.105 $ & $  903.25 $ & $  0.106x10^{16} $ & $ 1.867 $ & $ 2.235 $ \\
$ 23\,\, 14\,\, 58.7  $ &$ +14\,\,  5\,\, 24.4 $ & $ 0.083 $ & $  517.16 $ & $  0.212x10^{15} $ & $ 1.135 $ & $ 1.280 $ \\
$ 10\,\, 11\,\, 11.4  $ &$ + 8\,\, 41\,\, 31.9 $ & $ 0.097 $ & $  385.05 $ & $  0.140x10^{15} $ & $ 1.357 $ & $ 0.953 $ \\
$ 15\,\, 27\,\,  0.2  $ &$ +29\,\, 41\,\, 33.0 $ & $ 0.113 $ & $  689.54 $ & $  0.512x10^{15} $ & $ 1.543 $ & $ 1.706 $ \\
$ 15\,\, 51\,\, 37.4  $ &$ +45\,\, 33\,\, 16.2 $ & $ 0.124 $ & $  693.77 $ & $  0.441x10^{15} $ & $ 1.314 $ & $ 1.717 $ \\
$ 11\,\, 19\,\, 49.4  $ &$ +55\,\, 15\,\, 37.8 $ & $ 0.106 $ & $  537.54 $ & $  0.251x10^{15} $ & $ 1.245 $ & $ 1.330 $ \\
$  8\,\,  6\,\, 33.0  $ &$ +29\,\, 28\,\, 55.6 $ & $ 0.128 $ & $  458.29 $ & $  0.204x10^{15} $ & $ 1.390 $ & $ 1.134 $ \\
$  8\,\, 20\,\, 55.8  $ &$ + 7\,\, 52\,\, 12.7 $ & $ 0.110 $ & $  558.19 $ & $  0.287x10^{15} $ & $ 1.321 $ & $ 1.381 $ \\
$ 14\,\, 14\,\,  9.8  $ &$ - 0\,\,  8\,\, 26.9 $ & $ 0.139 $ & $ 1091.89 $ & $  0.190x10^{16} $ & $ 2.280 $ & $ 2.702 $ \\
$ 11\,\, 21\,\, 53.8  $ &$ + 0\,\, 41\,\, 55.0 $ & $ 0.102 $ & $  654.86 $ & $  0.422x10^{15} $ & $ 1.410 $ & $ 1.620 $ \\
$ 11\,\, 20\,\, 19.8  $ &$ +47\,\,  4\,\, 54.5 $ & $ 0.112 $ & $  572.08 $ & $  0.359x10^{15} $ & $ 1.570 $ & $ 1.416 $ \\
$ 10\,\, 48\,\, 58.7  $ &$ +54\,\, 53\,\, 22.9 $ & $ 0.145 $ & $  672.72 $ & $  0.557x10^{15} $ & $ 1.765 $ & $ 1.665 $ \\
$ 14\,\, 59\,\,  8.5  $ &$ +47\,\, 19\,\, 43.0 $ & $ 0.089 $ & $  677.50 $ & $  0.660x10^{15} $ & $ 2.062 $ & $ 1.676 $ \\
$ 10\,\, 27\,\, 49.7  $ &$ + 3\,\, 40\,\, 43.7 $ & $ 0.074 $ & $  819.40 $ & $  0.960x10^{15} $ & $ 2.051 $ & $ 2.027 $ \\
$ 21\,\, 29\,\, 58.2  $ &$ - 0\,\, 19\,\, 47.6 $ & $ 0.135 $ & $  323.62 $ & $  0.126x10^{15} $ & $ 1.721 $ & $ 0.801 $ \\
\hline
\end{longtable}
}
\normalsize
\end{appendix}


\end{document}